\documentclass[]{aa}
\usepackage{tabularx}
\usepackage{graphicx}
\usepackage{natbib}
\usepackage{txfonts}
\usepackage{longtable}
\usepackage{multirow}
\usepackage{lscape}
\usepackage{amssymb}
\usepackage{color}
\usepackage{xspace}

  \newcommand{\cmthree}{\;{\rm cm^{-3}}}
\newcommand{\hers}{{\it  Herschel}}

\newcommand{\spitz}{{\it   Spitzer}}

   \newcommand{\mic}{$\mu$m}
  \newcommand{\msol}{M$_\odot$}
  \newcommand{\lsol}{L$_\odot$}
  \newcommand{\zsol}{Z$_\odot$}
  
  \newcommand{\xco}{$X_{\rm CO}$}
  
  \newcommand{\smallsub}[1]{{\mbox{{\tiny #1}}}}

  \newcommand{\Av}{A$_\smallsub{V}$}

  \newcommand{\Nh}{N$_\smallsub{H}$}
   
  \newcommand{\n}{n$_\smallsub{H}$} 
  \newcommand{\den}{cm$^{-3}$}
  \newcommand{\go}{$G_{\rm 0}$}
  \newcommand{\co}{CO$\,$(1-0)}
   \newcommand{\hmol}{H$_2$}
  \newcommand{\hi}{H$\,${\sc i}}
  \newcommand{\hplus}{H$^+$}
  \newcommand{\hatom}{H$^0$}
  \newcommand{\hii}{H$\,${\sc ii}}
  \newcommand{\cplus}{C$^+$}
   \newcommand{\cone} {C$^{\rm 0}$}
  \newcommand{\cii}{[C$\,${\sc ii}]}
  \newcommand{\ci}{[C$\,${\sc i}]}
  \newcommand{\oiii}{[O$\,${\sc iii}]}
 \newcommand{\oi}{[O$\,${\sc i}]}
 \newcommand{\nii}{[N$\,${\sc ii}]}
 \newcommand{\niii}{[N$\,${\sc iii}]}

  \newcommand{\ciiline}{\cii$\lambda 158$\mic}
   
   \newcommand{\cilineup}{\ci$\lambda 610$\mic}
  \newcommand{\oiiiline}{\oiii$\lambda 88$\mic}
  \newcommand{\oilinelo}{\oi$\lambda 63$\mic}
  \newcommand{\oilineup}{\oi$\lambda 145$\mic}
  \newcommand{\niiline}{\nii$\lambda 122$\mic}
   \newcommand{\niilineup}{\nii$\lambda 205$\mic}
  \newcommand{\niiiline}{\niii$\lambda 57$\mic}

\newcommand{\ciico}{L$_\mathrm{[C\,{\sc II}]}$/L$_\mathrm{CO(1-0)}$}
\newcommand{\ciifir}{L$_\mathrm{[C\,{\sc II}]}$/L$_\mathrm{FIR}$}
\newcommand{\ciioifir}{L$_\mathrm{[C\,{\sc II}]+ [O\,{\sc I}]}$/L$_\mathrm{FIR}$}
\newcommand{\ciioi}{L$_\mathrm{[C\,{\sc II}]}$/L$_\mathrm{[O\,{\sc I}]}$}
\newcommand{\ciitir}{L$_\mathrm{[C\,{\sc II}]}$/L$_\mathrm{TIR}$}

\newcommand{\lcii}{L$_\mathrm{[C\,{\sc II}]}$}
\newcommand{\lci}{L$_\mathrm{[C\,{\sc I}]}$}
\newcommand{\lco}{L$_\mathrm{CO}$}
\newcommand{\lfir}{L$_\mathrm{FIR}$}
\newcommand{\loi}{L$_\mathrm{[O\,{\sc I}]}$}
\newcommand{\ltir}{L$_\mathrm{TIR}$}

\newcommand{\iizw}{\object{II$\;$Zw$\;$40}}
\newcommand{\mhmol}{M$_\mathrm{\tiny H_{2}}$}
\newcommand{\mhmolnoco}{M(H$_\mathrm{2}$)$_{\rm dark}$}
\newcommand{\mhmolco}{M(H$_\mathrm{2}$)$_{\rm CO}$}
\newcommand{\mhmoltot}{M(H$_\mathrm{2}$)$_{\rm total}$}

\def\revised{}

\title{Tracing the total molecular gas in galaxies: [CII] and the CO-dark gas}
 
\begin{document}

   \author{S. C. Madden
          \inst{1}
          \and
          D. Cormier
          \inst{1}
           \and
            S. Hony
          \inst{2} 
           \and
           V. Lebouteiller
           \inst{1}
           \and
           N. Abel
           \inst{3}
           \and
           M. Galametz
           \inst{1}
           \and
           I. De Looze
           \inst{4,5}  
           \and
           M. Chevance
           \inst{6}
           \and
           F. L. Polles
           \inst{7}
           \and
           M.-Y. Lee
           \inst{8}
           \and 
           F. Galliano
           \inst{1}
           \and
           A. Lambert-Huyghe
           \inst{1}
           \and
           D. Hu
           \inst{1}
           \and 
           L. Ramambason
           \inst{1}
          }

   \institute{AIM, CEA, CNRS, Universit\'e Paris-Saclay, Universit\'e Paris Diderot, Sorbonne Paris Cit\'e, 91191 Gif-sur-Yvette, France \\
              \email{suzanne.madden@cea.fr} \\
         \and
         Institut f\"ur theoretische Astrophysik, Zentrum f\"ur Astronomie der Universit\"at Heidelberg, Albert-Ueberle Str. 2,
69120 Heidelberg, Germany\\
         \and
         University of Cincinnati, Clermont College.  4200 Clermont College Drive, Batavia, OH, 45103 USA \\
 	\and
	Sterrenkundig Observatorium, Ghent University, Krijgslaan 281 S9, B-9000 Gent, Belgium \\
	\and
	Department of Physics and Astronomy, University College London, Gower Street, London WC1E 6BT, UK\\
	\and
	Astronomisches Rechen-Institut, Zentrum f\"ur Astronomie der Universit\"at Heidelberg, M\"onchhofstrasse 12-14, 69120 Heidelberg, Germany\\
	\and
	LERMA, Observatoire de Paris, PSL Research University, CNRS, Sorbonne Universit\'e, 75014 Paris, France\\
	\and
	Korea Astronomy and Space Science Institute, 776 Daedeokdae-ro, 34055, Daejeon, Republic of Korea\
             }

\date{Received July 6, 2020; Accepted August 14, 2020}
 
 \abstract 
 {Molecular gas is a necessary fuel for star formation. The \co\ transition is often used to deduce the total molecular hydrogen, but is challenging to detect in low metallicity galaxies, in spite of the star formation taking place. In contrast, the \ciiline\ is relatively bright, highlighting a potentially important reservoir of \hmol\ that is not traced by \co, but residing in the \cplus\ - emitting regions. } 
{Here we aim to explore a method to quantify the total \hmol\ mass (\mhmol) in galaxies and learn what parameters control the CO-dark reservoir.}
{We present Cloudy grids of density, radiation field and metallicity in terms of observed quantities, such as \oi, \ci, \co, \cii\ and \ltir\ and the total \mhmol.  We provide recipes based on these models to derive total \mhmol mass estimates from observations. We apply the models to the \hers\ Dwarf Galaxy Survey, extracting the total \mhmol\ for each galaxy and compare this to the \hmol\ determined from the observed \co\ line. This allows us to quantify the reservoir of \hmol\ that is CO-dark and traced by the \ciiline.}
{We demonstrate that while the \hmol\ traced by \co\ can be negligible, the \ciiline\ can trace the total \hmol. We find 70\% to 100 \% of the total \hmol\ mass is not traced by \co\ in the dwarf galaxies, but is well-traced by \ciiline. The CO-dark gas mass fraction correlates with the observed \ciico\ ratio. A conversion factor for \ciiline\ to total \hmol\ and a new CO-to-total-\mhmol\ conversion factor, as a function of metallicity, is presented.} 
 {While low metallicity galaxies may have a feeble molecular reservoir as surmised from CO observations, the presence of an important reservoir of molecular gas, not detected by CO, can exist. We suggest a general recipe to quantify the total mass of \hmol\ in galaxies, taking into account the CO and \cii\ observations. Accounting for this CO-dark \hmol\ gas, we find that the star forming dwarf galaxies now fall on the Schmidt-Kennicutt relation. Their star-forming efficiency is rather normal, since the reservoir from which they form stars is now more massive when introducing the \cii\ measures of the total \hmol, compared to the little amount of \hmol\ in the CO-emitting region.  }

   \keywords{galaxies: dwarf -- galaxies: ISM -- infrared: ISM -- photon-dominated region (PDR) -- HII regions 
               }
  \maketitle 
  %

\section{Introduction}
\label{intro} 
Our usual view of star formation posits the molecular gas reservoir as a necessary ingredient, the most abundant molecule being \hmol. This concept is borne out through testimonies of the first stages of star formation associated with and within molecular clouds and numerous observational studies showing the observed correlation of star formation with indirect tracers of \hmol\  reflected in the Schmidt-Kennicutt relationship \citep[e.g.][]{kennicutt98b,kennicutt07,bigiel08,leroy08,genzel12,kennicutt12,leroy13, kumari20}.  Witnessing the emission of \hmol\ associated with the bulk of the molecular clouds directly, however, is not feasible. Indeed, \hmol\ emits weakly in molecular clouds due to the lack of permanent dipole moment and the high temperatures necessary to excite even the lowest rotational transitions (the 2 lowest transitions have upper level energies, $h\nu/k$, of 510 K and 1015 K). Thus, \hmol\ observations can directly trace only a relatively small budget (15\% to 30\%) of warmer ($\sim$100K) molecular gas in galaxies \citep[e.g.][]{roussel07,togi16}, but not the larger reservoir of cold ( $\sim$10-20 K) molecular gas normally associated with star formation.   

In spite of 4 orders of magnitude lower abundance of CO compared to \hmol, most studies rely on CO rotational transitions to quantify the properties of the \hmol\ reservoir in galaxies.
The low excitation temperature of \co\ ($\sim 5$ K), along with its low critical density ($n_{\rm crit}$) for collisional excitation ($\sim 10^{3} \cmthree$), make it a relatively strong, easily-excited millimeter (mm) emission line to trace the cold gas in star-forming galaxies. Studies within our Galaxy have long ago established a convenient recipe to convert the observed \co\ emission line to \hmol\ gas mass, that is, the \xco\ conversion factor \citep[see][for historical and theoretical development]{bolatto13}. Likewise, galaxies in our local universe have also routinely relied on CO observations to quantify the \hmol\ reservoir \citep[e.g.][]{leroy11,bigiel11,schruba12,cormier14,cormier16,saintonge17}. CO is also commonly used to probe the  \hmol\ and star formation activity in the high-z (z$\sim$ 1 - 2) universe \citep[e.g.][]{tacconi10,daddi10a,combes13,carilli13,walter14,daddi15,genzel15,kamenetzky17,tacconi18,pavesi18, tacconi20}, although for higher-z sources, higher J CO lines may become more readily available, but less straightforward to relate to the bulk of the \hmol\ mass \citep[e.g.][]{papadopoulos08,gallerani14,kamenetzky18, vallini18,dessauges20}. Well-known caveats can hamper accurate total \hmol\ gas mass determination in galaxies based on observed CO emission only, including those associated with assumptions on CO excitation properties, dynamical effects related to \xco, presence of ensembles of molecular clouds along the lines of sight in galaxies and low filling factor and abundance of CO compared to \hmol, as in low metallicity environments \citep[e.g.][]{maloney88,glover11,shetty11,narayanan12b,pineda14,clark15,bisbas17}.

\begin{figure*}
 \begin{center}
  \includegraphics[width=14.0cm]{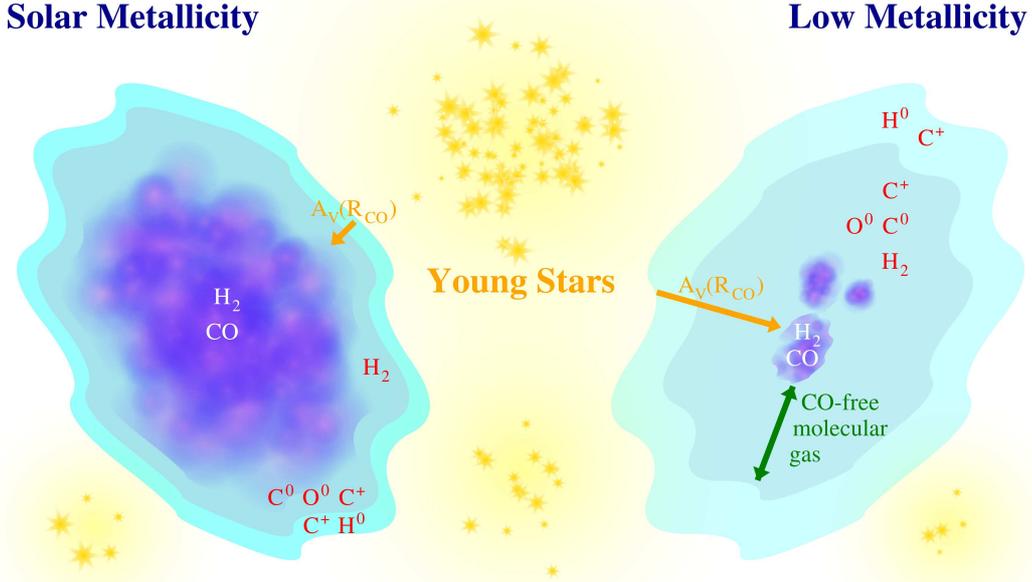}
  \caption{Comparison of a solar metallicity (metal-rich) molecular cloud and a low-metallicity cloud impacted by the UV photons of nearby star clusters. The decrease in dust shielding, in the case of the low metallicity cloud, leads to further photodissociation of the molecular gas, leaving a layer of self-shielded \hmol\ outside of the small CO-emitting cores. This CO-dark \hmol\ can, in principle, be traced by \cii\ or \ci.}
  \label{cloud}
 \end{center}  
 \end{figure*}

CO, however, can fail altogether to trace the full \hmol\ reservoir, very importantly, in low-metallicity galaxies. Lower dust abundance allows the far ultra-violet (FUV) photons to permeate deeper into molecular clouds compared to more metal-rich clouds, photodissociating CO molecules, leaving a larger \cplus-emitting envelope surrounding a small CO core. \hmol, on the other hand, photodissociates via absorption of Lyman-Werner band photons, which, for moderate \Av, can become optically thick, allowing \hmol\ to become self-shielded from photodissociation \citep[e.g.][]{gnedin14}, leaving a potentially significant reservoir of \hmol\ existing outside of the CO-emitting core, within the \cplus-emitting envelope (Fig.~\ref{cloud}). The presence of this CO-dark molecular gas \citep[e.g.][]{roellig06, wolfire10, glover12}, requires other means to trace this molecular gas reservoir that is not probed by CO. The total \hmol\ mass, then, is the CO-dark gas mass plus the \hmol\ within the CO-emitting core.  Efficient star formation in this CO-dark gas has been shown theoretically to be possible \citep[e.g.][]{krumholz11, glover12b}.

More recently, observations of the submillimetre (submm) transitions  of \ci\ have been used to trace the total \hmol\ gas mass in galaxies at high and low-z, especially due to the advent of the spectroscopic capabilities of SPIRE \citep{griffin10} on the {\it Herschel Space Observatory} \citep{pilbratt10} and with ALMA \citep{popping17,andreani18,jiao19,nesvadba19,bourne19,valentino20,dessauges20}, while theory and simulations \citep{papadopoulos04,offner14,tomassetti14,glover16,bothwell17,li18, heintz20} as well, deem \ci\ to be a viable tracer of CO-dark \hmol. Likewise, ALMA has opened the window to numerous detections of \ciiline\ in the high-z universe, making \cii, which is more luminous than \ci, a popular tracer of total \hmol\ in galaxies \cite[e.g.][]{zanella18,bethermin20,schaerer20,tacconi20}. Dust mass estimates via monochromatic or full spectral energy distribution (SED) studies have also been used to quantify the total \hmol\ gas mass in the nearby and distant universe \citep[e.g.][]{dacunha13,groves15} (more background on the different approaches to uncover CO-dark gas is presented in the following Section \ref{co_dark_studies}).

Local and global environmental effects, including star formation properties and metallicity, require consideration as well, when choosing wisely the approach to relate total gas mass to star formation. Low metallicity environments present well-known issues in pinning down the molecular gas mass. Detecting CO in dwarf galaxies has been notoriously difficult \citep[e.g.][and references within]{leroy09, schruba12,leroy12,cormier14,hunt15,grossi16,amorin16}, leaving us with uncertainties in quantifying the total molecular gas reservoir. 

Star-forming dwarf galaxies often have super star clusters which have stellar surface densities greatly in excess of normal \hii\ regions and OB associations. Thus, the dearth of detectable CO raises questions concerning the fueling of such active star formation. Possible explanations may include: 1) CO is not a reliable tracer of the total molecular gas reservoir; 2) there is a real under-abundance of molecular gas indicative of an excessively  high star formation efficiency; 3)  \hmol\ has been mostly consumed by star formation; 4) \hmol\ has been completely dissociated in the aftermath of star formation; 5) HI is the important reservoir to host star formation.
 
These issues compound the difficulty in understanding the processes of both local and global star formation in low-metallicity environments. {\bf Developing reliable calibrators for the total \mhmol\ in low Z galaxies is a necessary step to be able to refine recipes for converting gas into stars under early universe conditions and thus further our understanding of how galaxies evolve}.  
Understanding the detailed physics and chemistry of the structure and properties of the gas that fuels star formation can help us discriminate strategies for accurate methods to get to the bulk of the molecular gas in galaxies.   
 
Since carbon is one of the most abundant species in galaxies, its predominantly molecular form, CO, and its neutral and ionised forms, \cone and \cplus, carry important roles in the cooling of the interstellar medium (ISM) of galaxies.
The locations of the transitions from ionised carbon, to neutral carbon to CO, which are within the photodissociation region (PDR)-molecular cloud structure, depend on numerous local physical conditions, including the FUV radiation field strength (\go)\footnote{1 \go\ = 1.6 $\times$ 10$^{-3}$ erg cm$^{-2}$ s$^{-1}$, integrated over the energy range 6 to 13.6 eV \citep{habing68}.}, hydrogen number density (\n), gas abundances, metallicity (Z), dust properties, etc.  Low-metallicity star-forming environments seem to harbor unusual PDR properties compared to their dustier counterparts, with notably bright \ciiline\ emission compared to the often faint \co\ emission, characteristics that can be attributed to the lower dust abundance along with the star formation activity and clumpy ISM \citep[e.g.][]{cormier14,chevance16,madden19}.   
  
In this study we determine a basic  strategy to quantify the total \hmol\ mass in galaxies and estimate the CO-dark molecular gas mass existing outside the CO-emitting core, for a range of Z, \n\ and \go. We start with the models of \cite{cormier15,cormier19}, which were generated using the spectral synthesis code, Cloudy \citep{ferland17}, to self-consistently model the mid-infrared (MIR) and far-infrared (FIR) fine structure cooling lines and the total infrared luminosity (\ltir) from \spitz\ and \hers. This approach allows us to constrain the photoionization and PDR region properties, quantifying the density and incident radiation field on the illuminated face of the PDR, \go. These models exploit the fact that there is a physical continuity from the PDR envelope, where the molecular gas is predominantly CO-dark, into the CO-emitting molecular cloud. This dark gas reservoir can be primarily traced by the \ciiline\ line while the \ciico\ can constrain the depth of the molecular cloud and the \ciitir\ and \ciioi\ constrain the \go\ and density at the illuminated face of the PDR (discussed in Section \ref{model_grids}). With this study, we determine a conversion factor to pin down the total \hmol, which will make use of the observed  \cii,  \oi,  \co\ and \ltir. 
Application of the models to the star-forming low-metallicity galaxies of the \hers\ Dwarf Galaxy Survey \citep[DGS;][]{madden13}, using the observed \cii\ and \co\ observations, allows us to extract the total \hmol\ mass and determine the fraction of the total \hmol\ reservoir that is not traced by CO observations, as a function of metallicity and other galactic parameters.
 
This paper is organised as follows. In Section \ref{co_dark_studies} we review the studies that have focused on uncovering the CO-dark reservoir in galaxies. The environments of the dwarf galaxies in the DGS are briefly summarised in Section \ref{DGS_summary}. In Section \ref{cii_co_fir} we bring in the observational context and motivation by comparing the \ciico\ and \ciifir\ \footnote{Note that for historical reasons, we use \lfir\ when comparing observations of the DGS to other observations collected in the literature over the years, originally presented with \lfir, such as Fig. \ref{cii_co_fir_fig}, adapting \lfir\ and \ltir\ as per \cite{remy13,remy15}. The , however, uses \ltir\ (3 to 1100 \mic), which is more commonly used today.} of the full sample of the DGS to other more metal-rich galaxies from the literature. In Section \ref{calculations} we present Cloudy model grids and discuss the effects of \go, \n\ and metallicity. In this section we walk through the steps to use the models to understand the mass budget of the carbon bearing species in the PDRs and to obtain the total \hmol. In Section \ref{example} we apply the models to the particular case of \iizw\ observations, as an example. Section \ref{dgs} discusses the model results applied to the DGS targets. In this section we extract the total \hmol\ for these galaxies and discuss the implication of the CO-dark gas reservoirs. We provide recipes to estimate the CO-dark \hmol\  and the total \hmol\ gas reservoir from \cii. Possible caveats and limitations are noted in Section \ref{caveats} and the results are summarised in Section \ref{conclusions}.

\section{CO dark gas studies}
\label{co_dark_studies}
Three decades ago, the FIR spectroscopic window became readily accessible from the {\it Kuiper Airborne Observatory} (KAO), and the {\it Infrared Space Observatory} (ISO). First estimates of the CO-dark \hmol\ reservoir obtained from \ciiline\ observations found that 10 to 100 times more \hmol\ was 'hidden' in the \cplus-emitting regions in dwarf galaxies compared to that inferred from \co\ alone \citep{poglitsch95, madden97}. These early estimates, for lack of access to additional complementary FIR diagnostics to carry out detailed , relied on a number of assumptions which have, until now, limited the use of the FIR lines to robustly quantify this CO-dark molecular gas in galaxies. Recent spatially-resolved studies of [CII] in Local Group galaxies using {\it Herschel} and the {\it Stratospheric Observatory for Infrared Astronomy} (SOFIA), have also confirmed a significant presence of CO-dark gas \citep[e.g.][]{fahrion17,requena-torres16,jameson18, lebouteiller19,chevance20}. Additionally, the MIR rotational transitions of \hmol\ from \spitz\ have been invoked to deduce a significant reservoir of \hmol\ in dwarf galaxies \citep{togi16}. Other molecules observed in absorption, such as OH and HCO$^+$, have also been used as tracers of dark molecular gas in our Galaxy \citep{lucas96,allen15,xu16,li18,nguyen18,liszt19}. However, the need for high enough sensitivity and resolution limits their use in other galaxies.
 
As $\gamma$ rays interact with hydrogen nuclei, they can trace the total interstellar gas. Comparison of $\gamma$ ray observations in our Galaxy with CO and \hatom\ has uncovered an important reservoir of neutral dark gas that is not  traced by CO or \hi\ \citep{grenier05,ackermann12, hayashi19}. While $\gamma$ ray emission, in principle, can be one of the most accurate methods possible to get at the total gas mass, its use to measure the total interstellar gas mass in other galaxies, is limited for now, because of the need for relatively high resolution observations. 

Comparison of \hers\ \cii\ velocity profiles to \hi\ emission \citep{langer10,pineda13,langer14, tang16} as well as observations comparing dust, \co\ and \hi\ \citep{planck11,reach17}, confirm that this CO-dark gas reservoir can be an important component in our Galaxy, and comparable to that traced by \co\ alone. Recent hydrodynamical simulations with radiative transfer and analytic theory also demonstrate the presence of this dark gas reservoir in the Galaxy, that should be traceable via \cii\ \citep{smith14,offner14,glover16,nordon16,gong18,franeck18,seifried20}, often associated with spiral arms in disk galaxies.  While optically-thick \hi\ can, in principle, be the source of the CO-dark gas, there is no compelling evidence that it is a strong contribution to the dark neutral gas \citep[e.g.][]{murray18}. Thus we focus here on the CO-dark \hmol\  gas, and the potential of \cplus, in particular, to quantify the \hmol\ reservoir.

Dust mass measurements with {\it Planck}, compared to the gas mass measurements, have uncovered a reservoir of dark gas in the Galaxy \citep{planck11,reach15}. Dust measurements have often provided easier means, from an observational strategy, to quantify the full reservoir of  gas mass of galaxies in the local universe as well as distant galaxies \citep[e.g.][]{magdis11,leroy11,magnelli12,eales12,bourne13,sandstrom13,groves15,scoville16,liang18,bertemes18}. This approach also poses its own fundamental issues \citep[e.g.][]{privon18}. The derived total molecular gas mass is a {\it difference} measure of two large quantities (\hi-derived atomic gas mass and dust-derived total gas mass), which can result in large uncertainties. Dust mass determination, which is commonly derived via the modelling of the SED, requires constraints that include submm observations and necessitates dust models with assumed optical properties as a function of wavelength, composition and size distributions. Then, turning total dust mass into total gas mass requires assumptions on dust-to-gas mass ratios which can carry large uncertainties depending on star formation history and metallicity and can vary by orders of magnitude, as shown by statistical studies of low metallicity galaxies \citep[e.g.][]{remy14,roman-duval14,galliano18}. 

One contributing factor to the difficulty in being conclusive on the relationship of gas, dust and star formation in low-metallicity conditions, in particular, is the fact that galaxies with full MIR to submm dust and gas modelling have been limited to mostly metal-rich environments and relatively higher star formation properties due to telescope sensitivity limitations.
 This impediment has been ameliorated with the broad wavelength coverage, the spatial resolution and sensitivity of the {\it Herschel}, allowing the accessibility of the gas and dust properties of low-metallicity galaxies. The DGS has compiled a large observational data base of 48 low-metallicity galaxies \citep{madden13}, motivated by the \hers\ PACS \citep{poglitsch10} and \hers\ SPIRE  55 to 500 \mic\  photometry and spectroscopy capabilities.  

\section{Dwarf Galaxy Survey - extreme properties in low metallicity environments}
\label{DGS_summary}
The DGS  targeted the most important FIR diagnostic tracers, \ciiline, \oilinelo, \oilineup, \oiiiline, \niiiline, \niiline\ and \niilineup\ \citep{cormier12,cormier15,cormier19} in a wide range of low-Z galaxies -  as low as $\sim$1/50 \zsol. Additionally, the DGS collected all of the FIR and submm photometry from \hers\ to investigate the dust properties of star-forming dwarf galaxies. The infrared SEDs exhibit distinct characteristics setting them apart from higher metallicity galaxies with different dust properties, generally exhibiting overall warmer dust than normal metallicity galaxies; an obvious paucity of PAH molecules and a striking non-linear drop in dust-to-gas mass ratios at lower metallicities (Z $<$ 0.1 \zsol) \citep[e.g.][]{galametz11, dale12, remy13, remy15}

\cii\ usually ranks foremost amongst the PDR cooling lines in normal metallicity galaxies, followed by the \oilinelo, with the \ciioifir\ often used as a proxy of the heating efficiency of the photoelectric effect \citep[e.g.][]{croxall12, lebouteiller12, lebouteiller19}.  \cite{cormier15} summarised the observed FIR fine structure lines in the DGS noting that the range of \ciioifir\ of low-metallicity galaxies is higher than that for galaxies in other surveys of mostly metal-rich galaxies \citep[e.g.][]{brauher08,croxall12,vanderlaan15,cigan16,smith17,lapham17, diazsantos17}, indicating relatively high photoelectric heating efficiency in the neutral gas. In star-forming dwarf galaxies, however, the \oiiiline\ line is the brightest FIR line, not the \ciiline\ line, as noted on full-galaxy scales \citep[e.g.][]{cormier15,cormier19} as well as resolved scales \citep{chevance16,jameson18,polles19}. In more quiescent metal-poor galaxies, however, the \ciiline\ line is brighter than the \oiiiline\ line \citep[e.g.][]{cigan16}. The predominance of  the large scale \oiiiline\ line emission, which requires an ionization energy of 35 eV, demonstrates the ease at which such hard photons can traverse the ISM on full galaxy scales in star-forming dwarf galaxies \citep{cormier10,cormier12,cormier15,cormier19}, highlighting the different nature and the porous structure of the ISM of low metallicity galaxies. Since the \oiii/\cii\ has been observed to also be similarly extreme in many high-z galaxies \citep[e.g.][]{laporte19,hashimoto19,tamura19,harikane20}, the results presented here, where we focus on low-Z star-forming galaxies, may eventually be relevant to the understanding of the total molecular gas content and structure in some high-z galaxies.
 
\begin{figure*}
 \begin{center}
  \includegraphics[width=12.0cm]{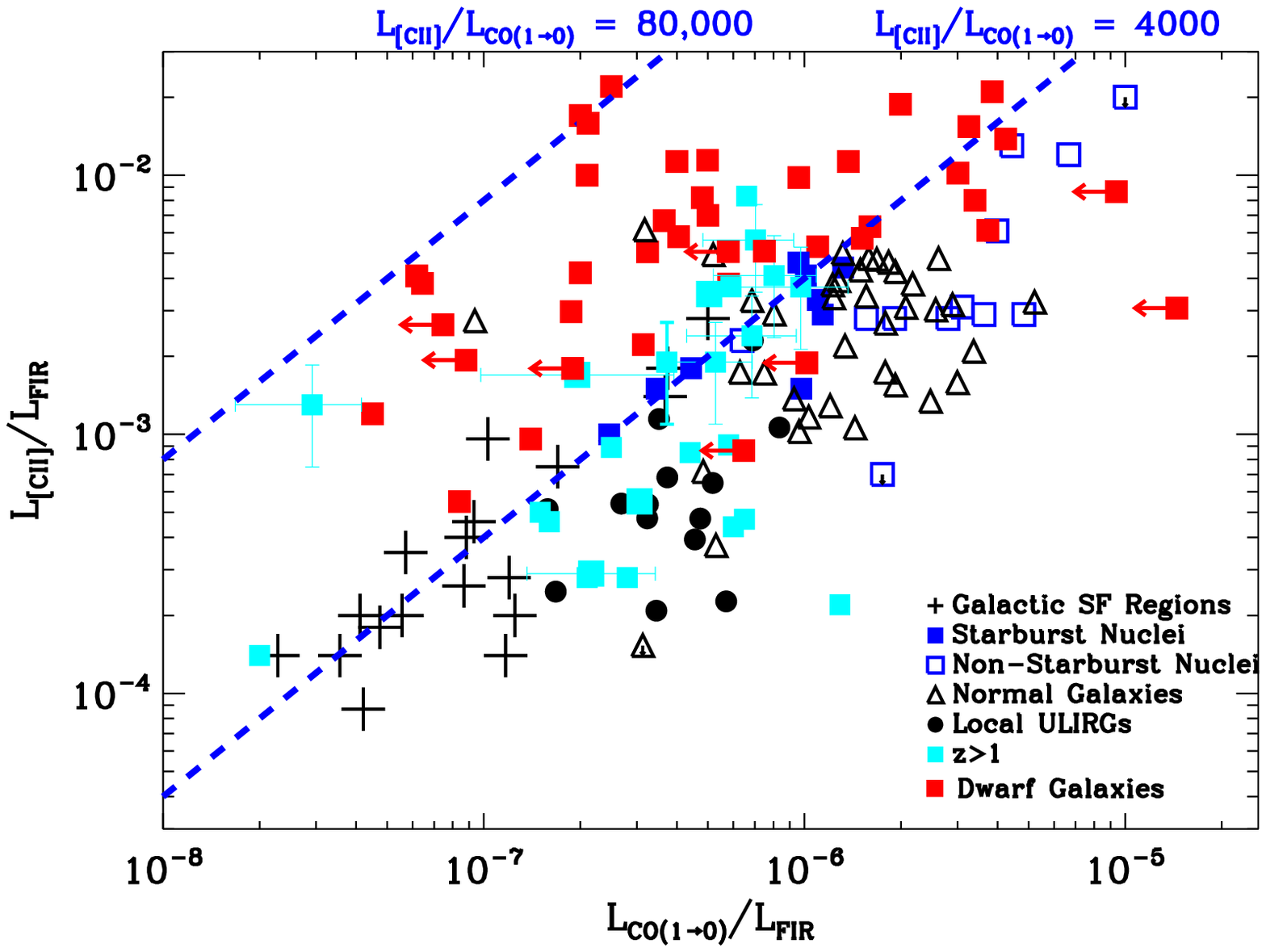}
  \caption{L$_{CO(1-0)}$/\lfir\ vs \ciifir\ observed in galaxies ranging widely in type, 
    metallicity and in star formation properties. This is updated from \cite{stacey91, madden00,  haileydunsheath10, stacey10} to include dwarf galaxies and more high redshift galaxies. The dwarf galaxies are from \cite{cormier10, cormier15} for the DGS; \cite{grossi16} for HeVICs dwarfs and \cite{smith97}. High redshift galaxies include those from \cite{haileydunsheath10, stacey10, gullberg15}. The black and blue symbols are data from the original figure of \cite{stacey91} for Galactic star-forming regions, starburst nuclei and non-starburst nuclei, ULIRGS \citep{luhman03} and normal galaxies \citep{malhotra01}. The dashed lines are lines of constant \ciico. Note the location of the low metallicity dwarf galaxies (red squares) which show extreme observed \ciico\ values.
 }
  \label{cii_co_fir_fig}
 \end{center}
  \end{figure*}

\section{\ciico\ in Galaxies}
\label{cii_co_fir}
Figure \ref{cii_co_fir_fig} shows the \co\ luminosity (\lco) and \cii\ luminosities (\lcii), normalised by FIR luminosity (\lfir), for a wide variety of environments, ranging from starburst galaxies and Galactic star forming regions, to less active, more quiescent "normal" galaxies, low metallicity star forming dwarf galaxies and some high-metallicity galaxies as well as some high-z galaxies. This figure, initially presented by \cite{stacey91} and subsequently used as a star-formation activity diagnostic, indicates that most normal and star forming galaxies are observed to have \ciico\ $\sim$ 1000 to 4000, with the more active starburst galaxies showing approximately a factor of 3 higher range of \ciico\ than the more quiescent galaxies \citep{stacey91,stacey10,haileydunsheath10}. The more active, dusty, star forming environments possess widespread PDRs exposed to intense FUV and the \ciico\ depends on the strength of the UV field and {\revised the shielding of the CO molecule against photodissociation due to \hmol\ and dust} \citep[e.g.][]{stacey91,stacey10,wolfire10,accurso17b}.  

It was already shown that \lcii\ was somewhat enhanced relative to \lfir\ and more remarkably so, relative to \lco\ in a few cases of star-forming low-metallicity galaxies compared to more metal-rich galaxies \citep{stacey91, poglitsch95, israel96, madden97, hunter01, madden11,madden19}.  
In recent, larger studies, \cite{cormier15,cormier19} have carried out detailed modelling of the DGS galaxies and compared their overall physical properties to those of more metal-rich galaxies attributing the enhanced \ciifir\ to the synergy of their decreased dust abundance and active star formation. The consequence is a high ionization parameter along with low densities resulting in a thick cloud and considerable geometric dilution of the UV radiation field. The effect over full galaxy scales is a low average ambient radiation field (\go) and relatively normal PDR gas densities (often of the order of 10$^3$ to 10$^4$ cm$^{-3}$), with relatively large \cplus\ layers. This scenario is also coherent with the \ciico\ observed in the low metallcity galaxies.

\begin{figure*}
\centering
\includegraphics[width=7.0cm]{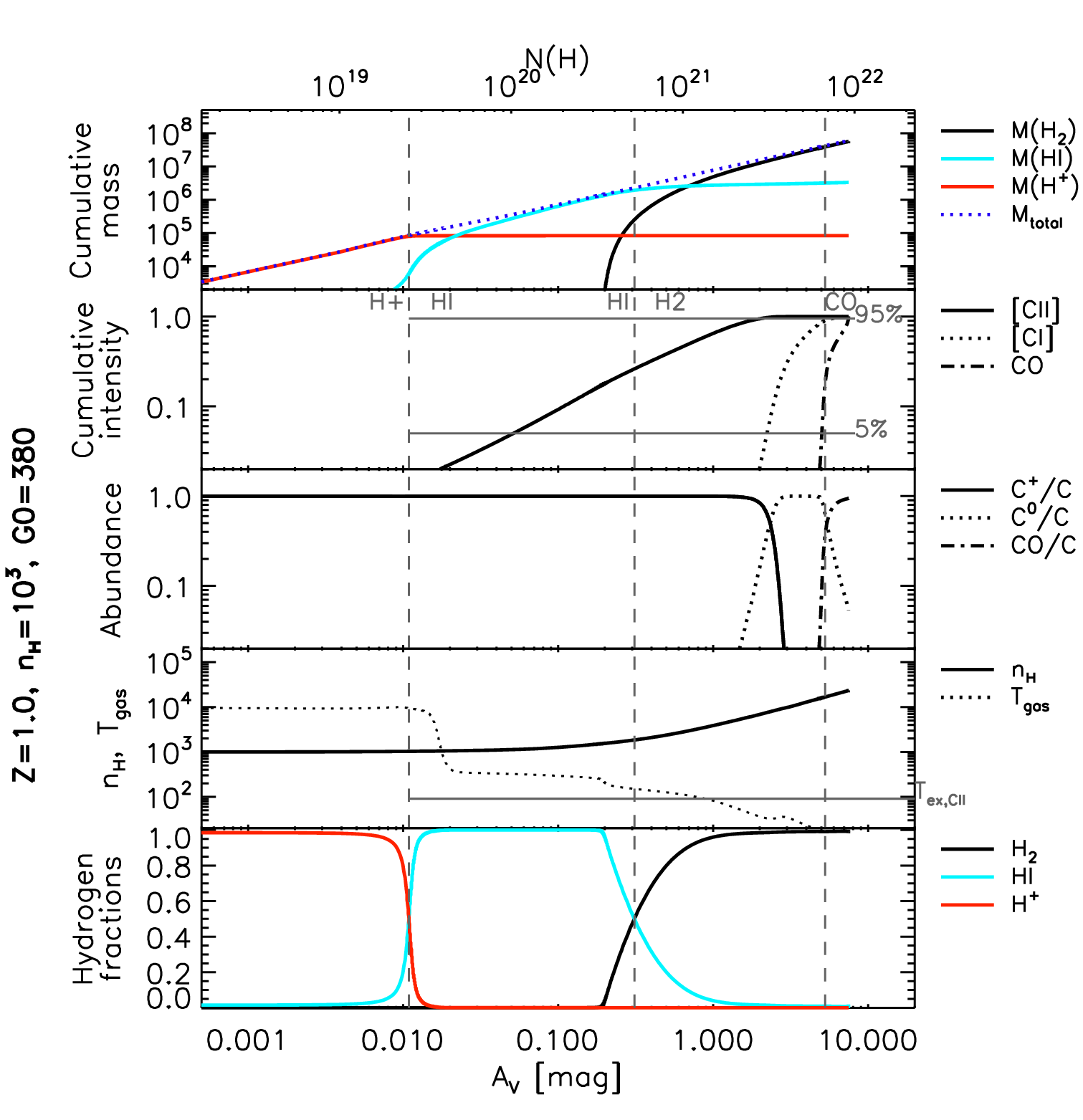}
\hspace{5mm}
\includegraphics[width=7.0cm]{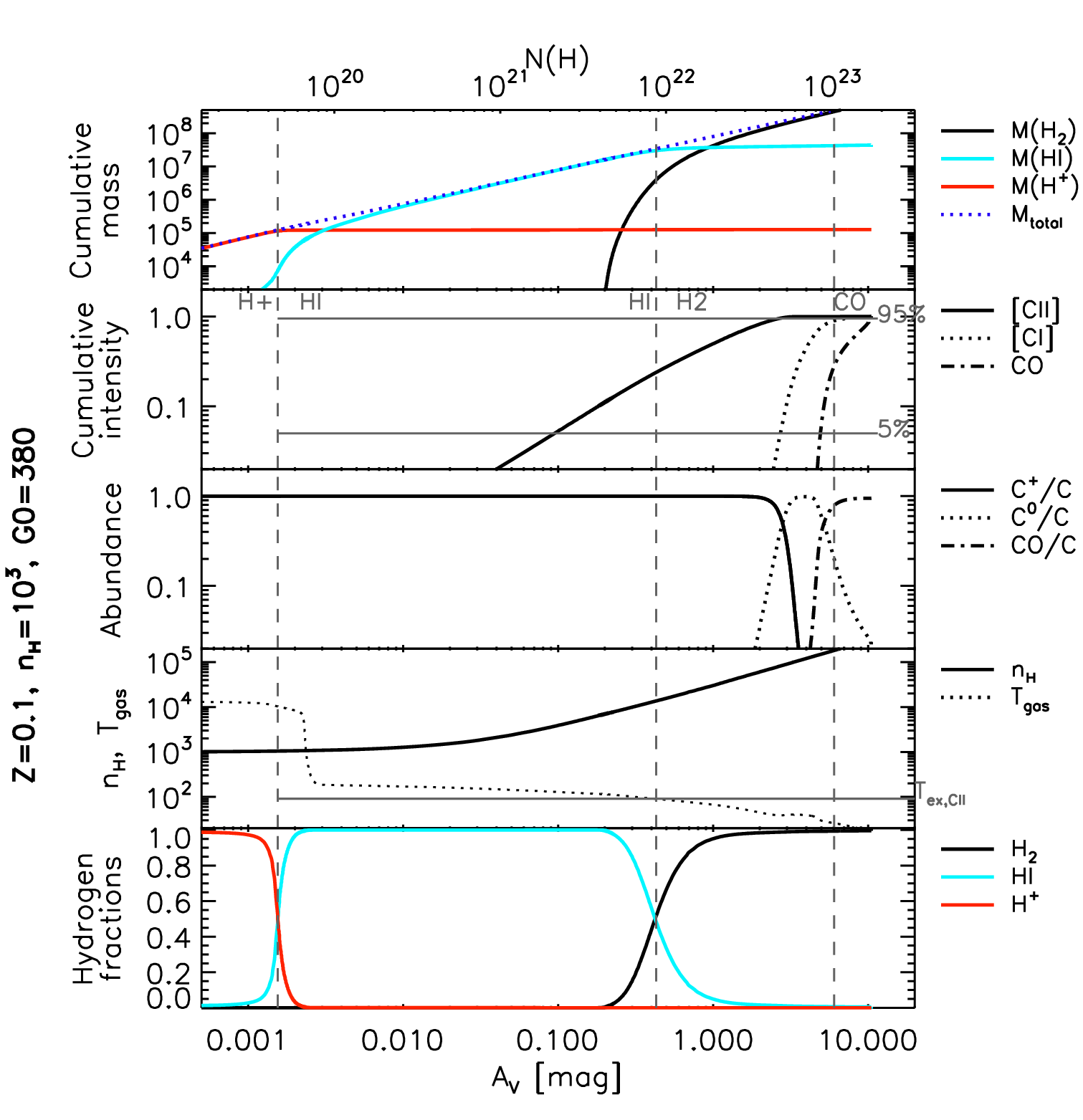}
\hspace{5mm}
\includegraphics[width=7.0cm]{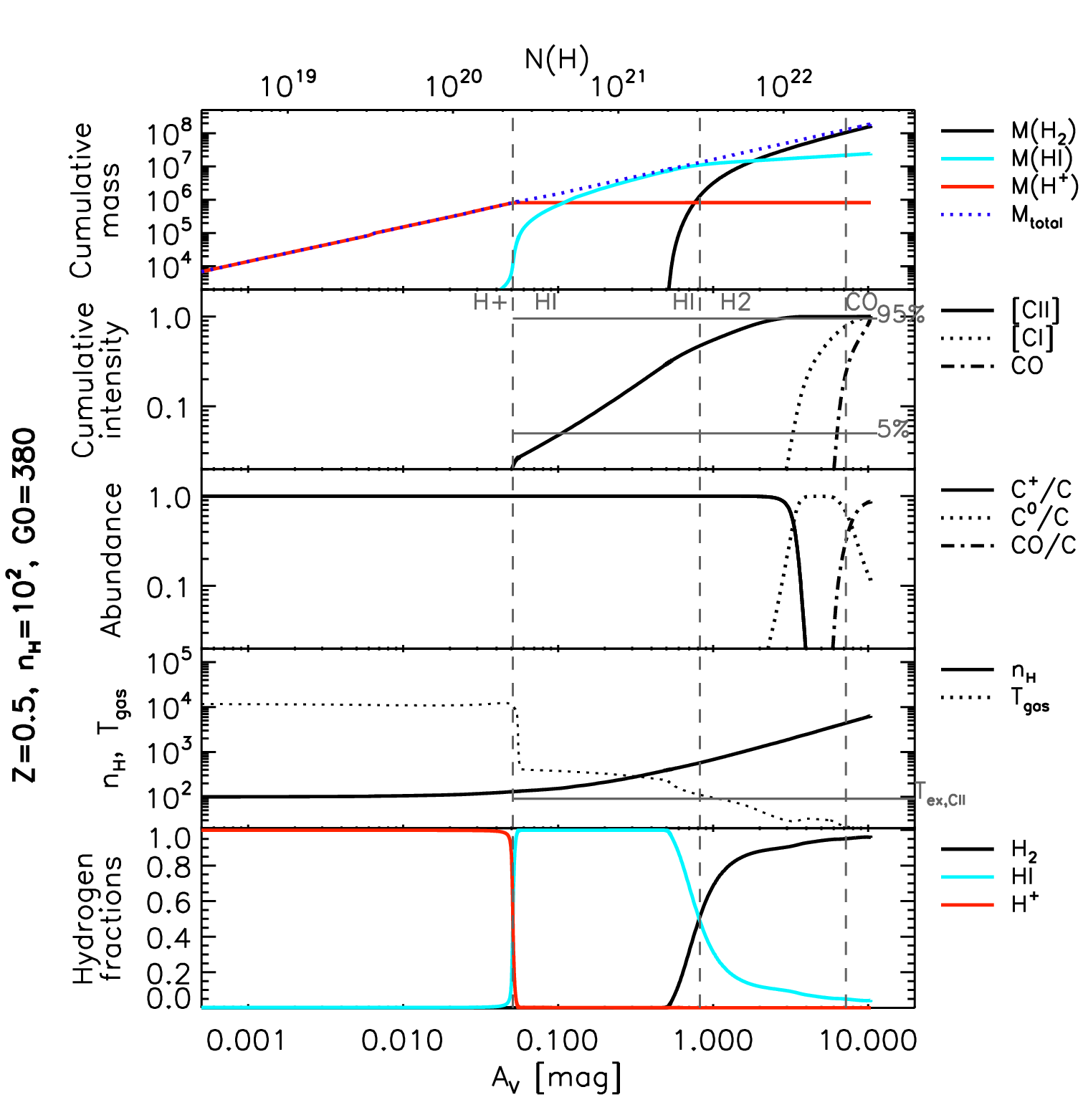}
\hspace{5mm}
\includegraphics[width=7.0cm]{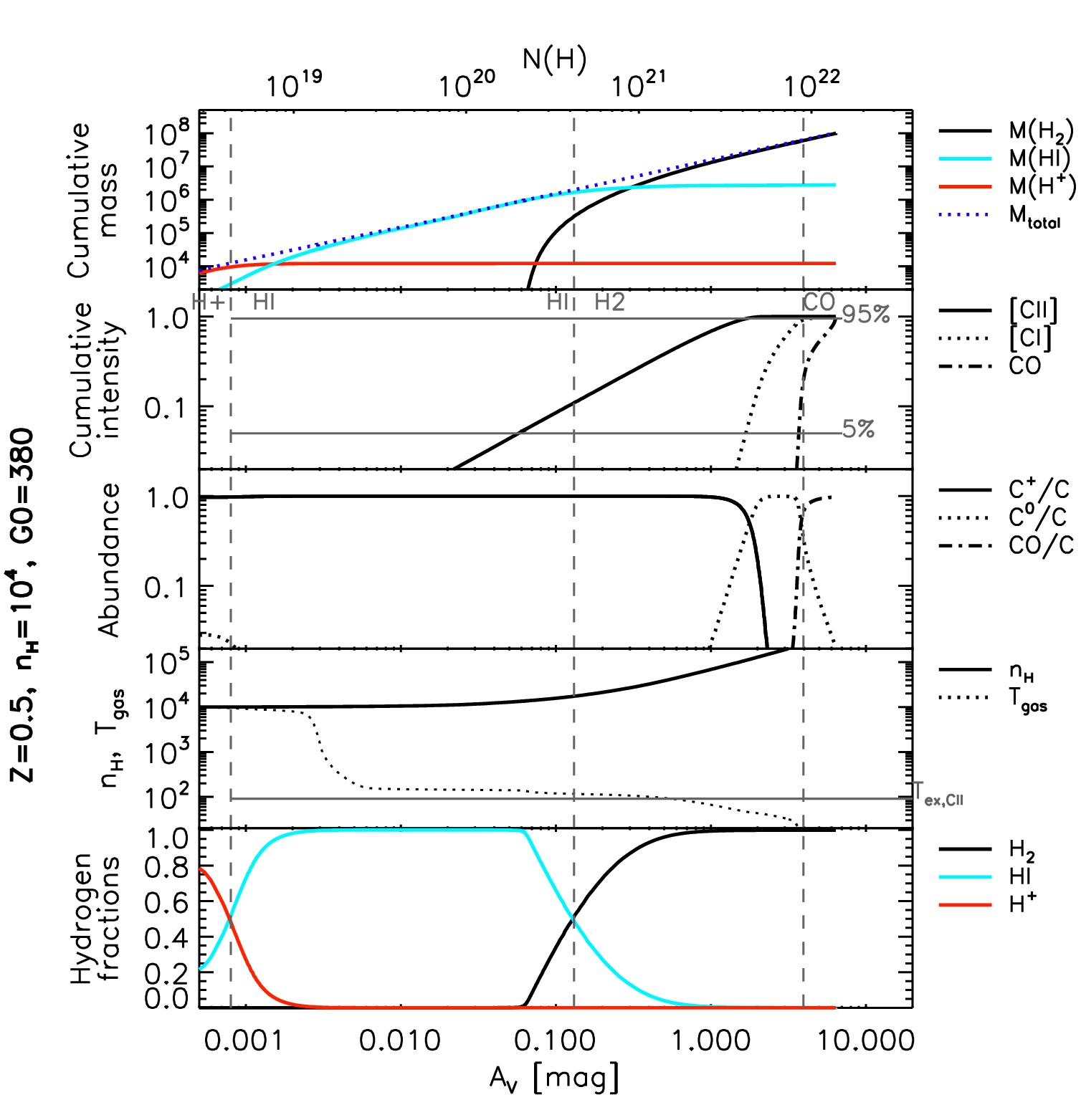}
\hspace{5mm}
\includegraphics[width=7.0cm]{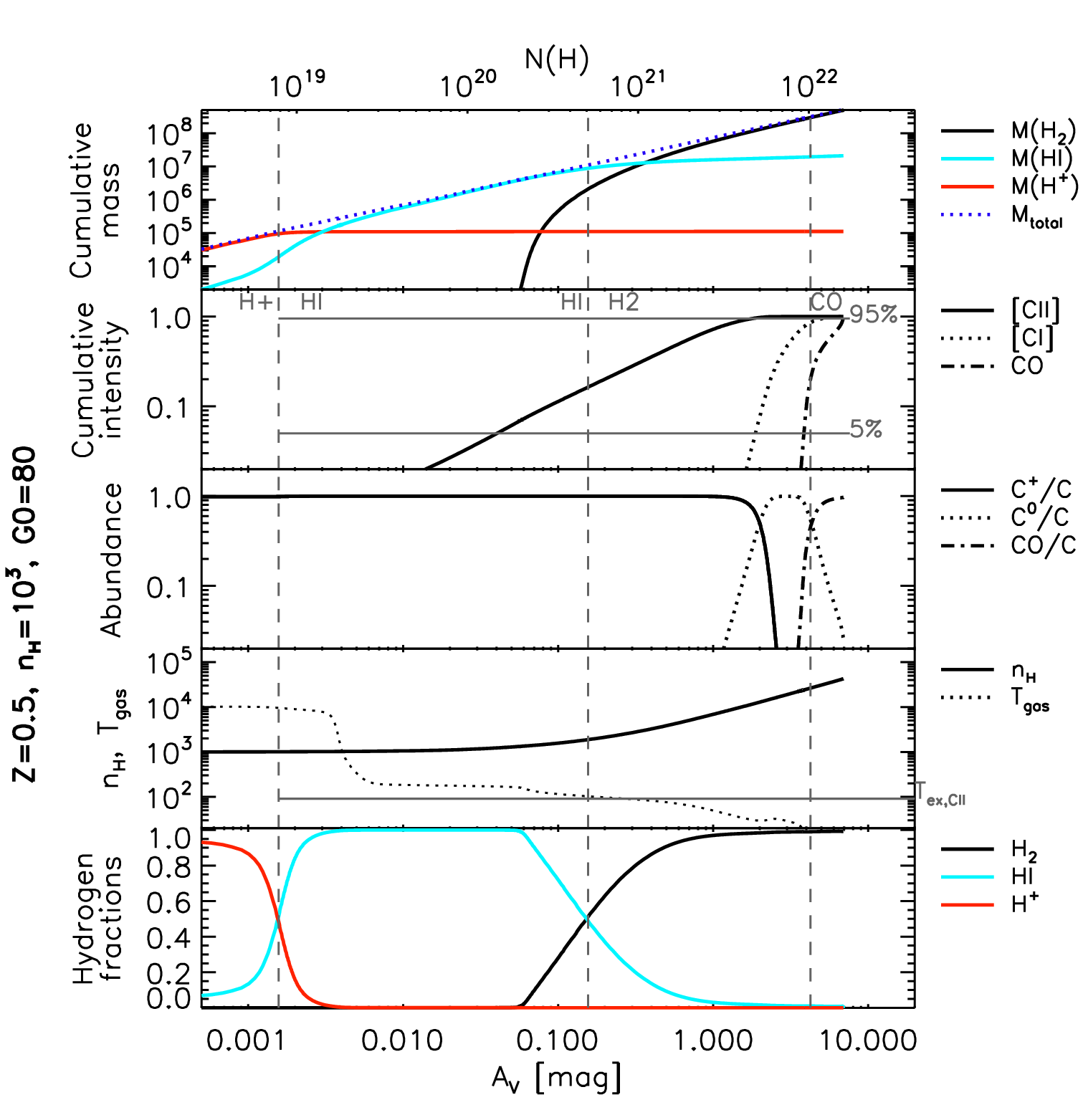}
\hspace{5mm}
\includegraphics[width=7.0cm]{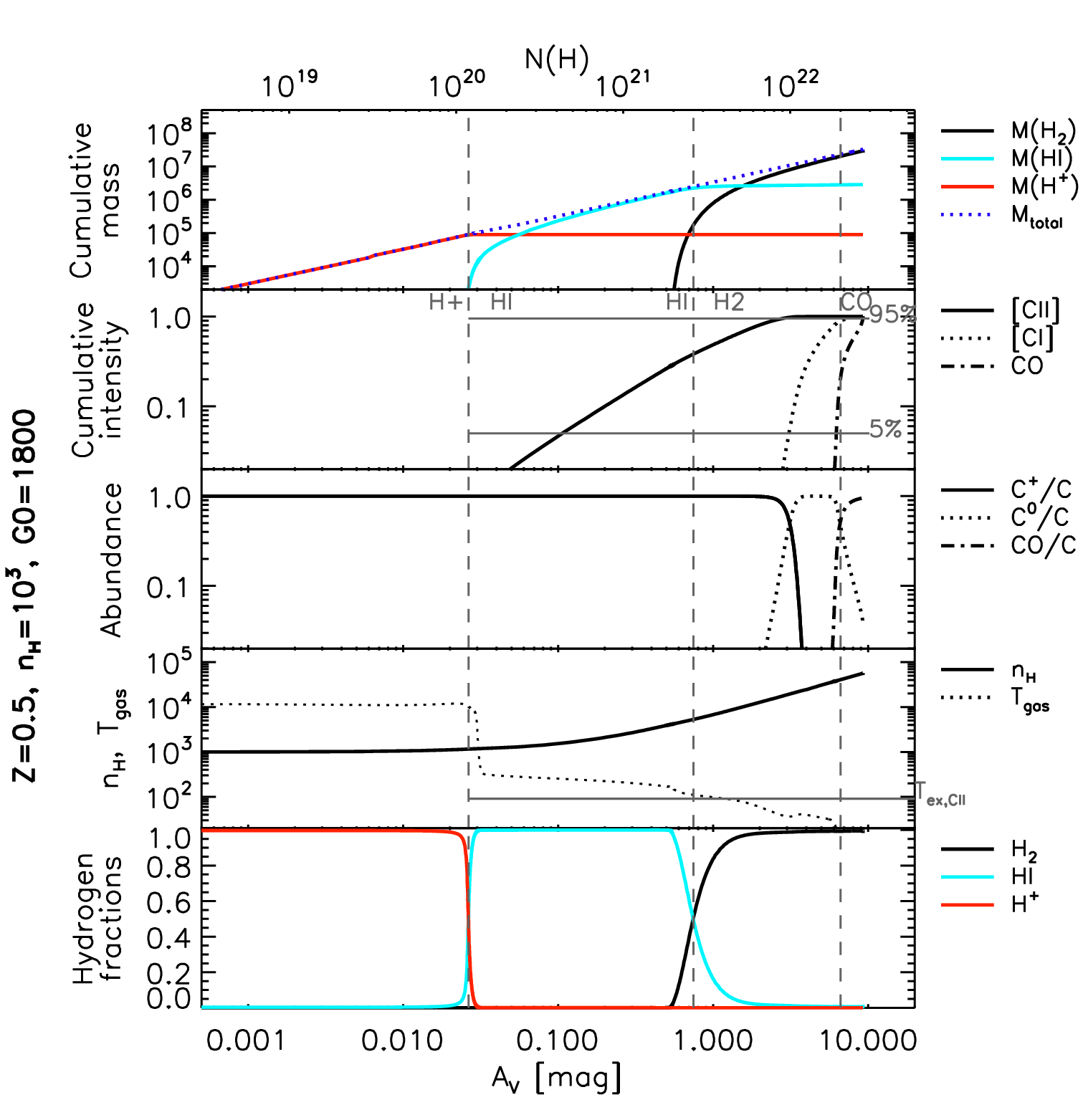} 
\hspace{5mm}
\caption{Evolution of the modeled gas parameters, Z, \n, \Nh, as a function of depth (\Av) for a solar-metallicity cloud ({\it top left}) and for a 0.1\zsol ({\it top right}), with a starting density of $10^3$\,\den\ and $r_{\rm in}$ of $10^{20.7}$\,cm,  \go\ = 380 in terms of the Habing field. {\it Center} two panels show the effect of density variations ({\it left}: \n\ = $10^2$\,\den; {\it right}: \n\ = $10^4$\,\den) for Z=0.5 \zsol\ with \go = 380.  {\it Bottom} two panels show the effect of \go\ variations ({\it left}: \go = 80;  {\it right}: \go\ =1800) for constant Z (0.5 \zsol) and constant \n\ ($10^3$\,\den). Each panel contains subpanels within, from top to bottom: cumulative mass (\msol) in the ionised, neutral atomic, and molecular phases; normalised cumulative intensity of the \ciiline, \cilineup\ and \co\ lines; abundance of \cplus, \cone, and CO relative to C; hydrogen density (\n) as a function of column density (\Nh) and \Av; fraction of hydrogen in the ionised, atomic, and molecular form. The vertical dashed lines indicate the depth of the main phase transitions: H$^+$ to H$^{\rm 0}$, H$^{\rm 0}$ to \hmol, CO(1-0) optical depth of 1. Masses indicated here should be adjusted for individual galaxies, scaling as \ltir(galaxy)/10$^{9}$\lsol, as the source luminosity of the model is 10$^9$ \lsol\ (see Appendix \ref{appendix}).
}
\label{phase_schematics}
\end{figure*}

As can be seen in Fig. \ref{cii_co_fir_fig}, star-forming dwarf galaxies can reach \ciico\ ratios an order of magnitude higher, or more (sometimes reaching 80,000) than their more metal-rich counterparts, on galaxy-wide scales. While  \co\ has been difficult to detect in star-forming low metallicity galaxies, thus shifting these galaxies well-off the Schmidt-Kennicutt relation \citep[e.g.][]{cormier14}, \cii, on the other hand,  has been shown to be an excellent star formation tracer over a wide range of galactic environments, including the dwarf galaxies \citep[e.g.][]{delooze11,delooze14,herrera_camus15}. Such relatively high \cii\ luminosities in star-forming dwarf galaxies, harboring little \co, may be indicative of a reservoir of CO-dark gas, which is one of the motivations for this study. 
 
\section{Using spectral synthesis codes to characterise the ISM physical conditions}
\label{calculations}
The strategy of this study is to first generate model grids that will help us explore how the properties of the CO-dark gas evolve as a function of local galaxy properties, such as Z, \n\ and \go, and then determine how observational parameters can be used to pin down the mass of the CO-dark gas.  We will show how \ciico\ can constrain \Av\ from the models and then the \ciitir\ can narrow down the values of \n\ and \go\ to finally quantify the total mass of \hmol\ and hence, the mass of \hmol\ that is not traced by CO, the CO-dark gas.  
 
 \begin{figure*}
\includegraphics[width=19.0cm]{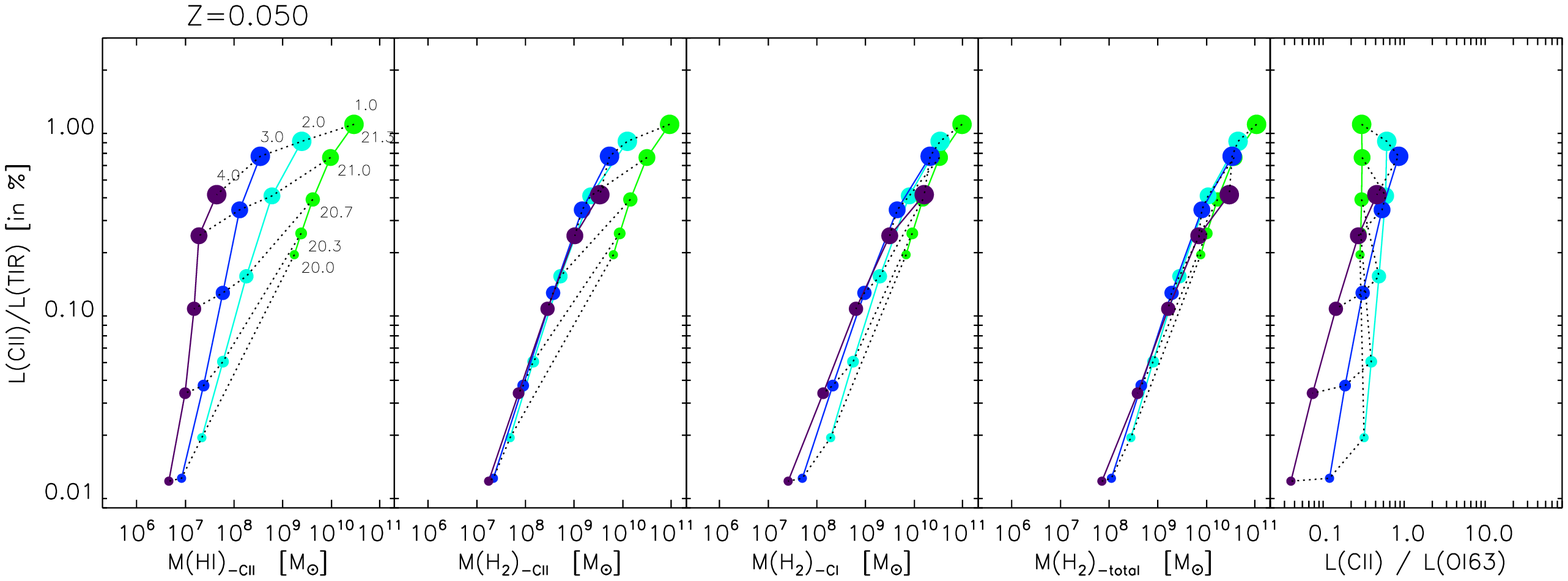}
\includegraphics[width=19.0cm]{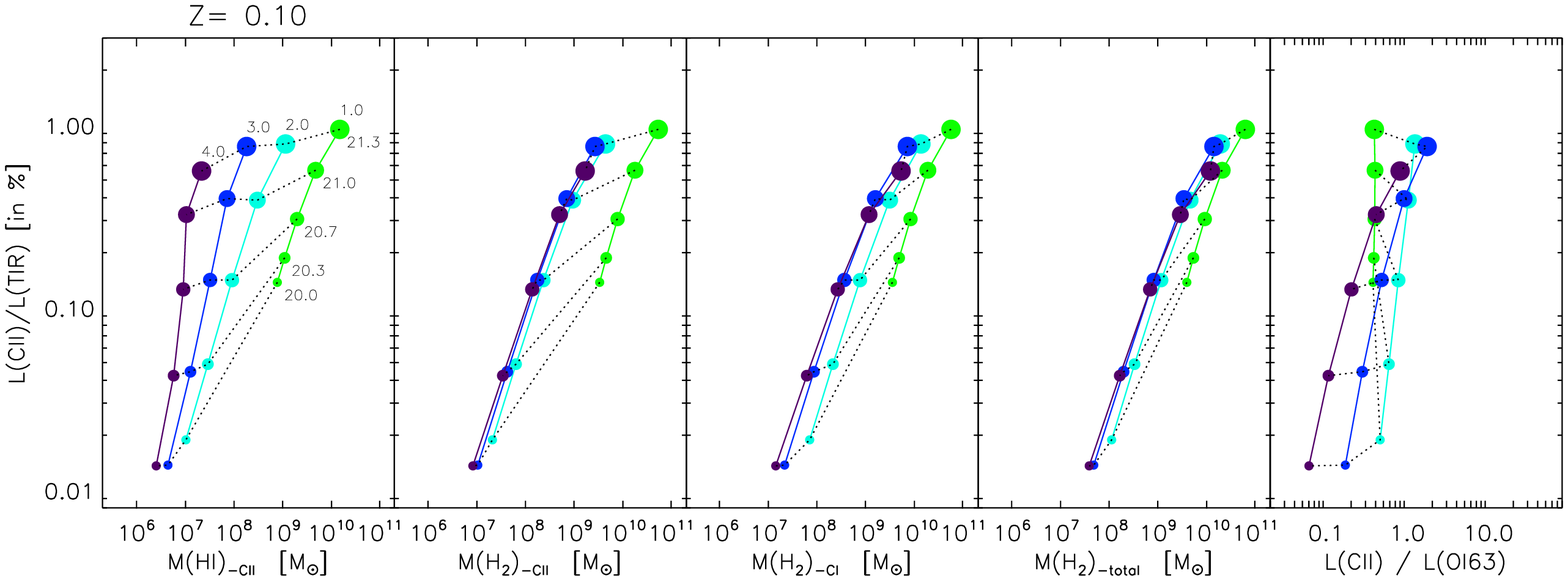}
\includegraphics[width=19.0cm]{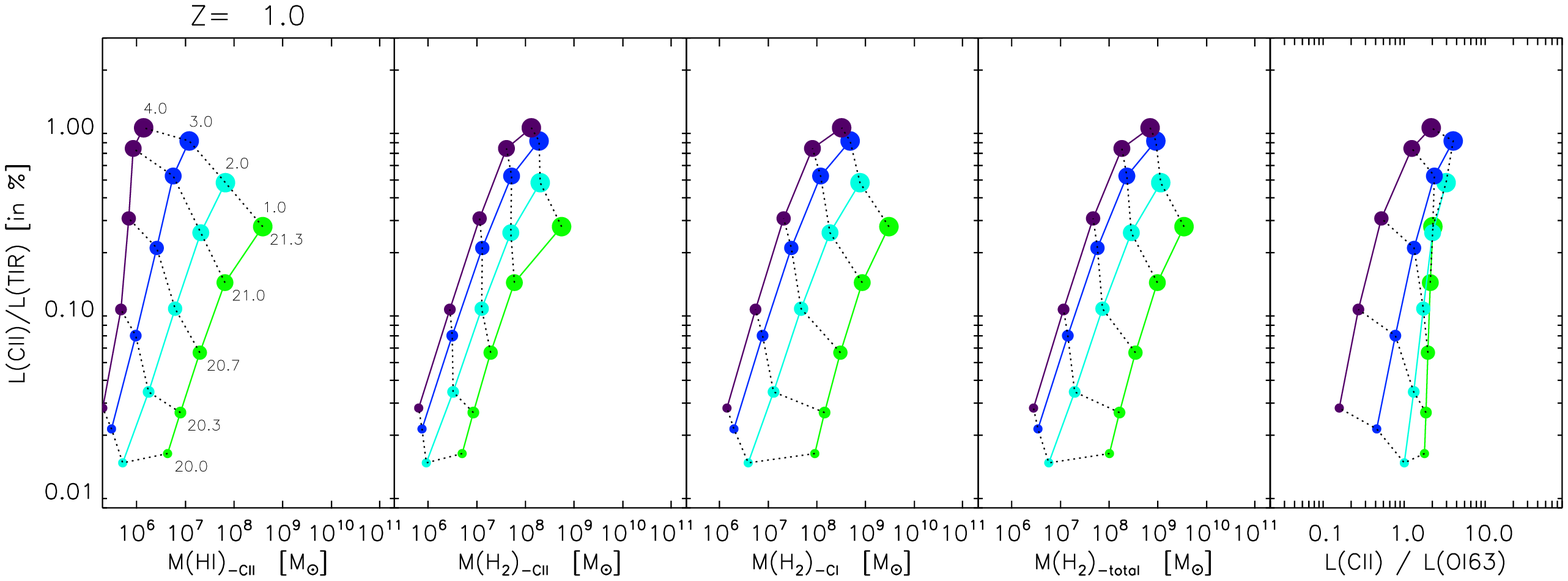}
\caption{Grids of Cloudy calculations:  Model \ciitir\ vs \hi\ and \hmol\  mass reservoirs in the \cplus, \cone-emitting regions, total \hmol\ mass, and \ciioi\ behavior (last column) in terms of metallicities of Z=0.05, 0.1, 1.0 \zsol\ ({\it from top to bottom}) and for a range of \go\ and \n, the initial hydrogen density.  The colour coding in each figure is log \n/cm$^{-3}$\ which increases from 1.0 (green) to 4.0  (purple). A range of \go\ is set by varying log r$_{in}$ /cm from 21.3 at the top right of the grids (large dots) to 20.0 at the bottom left of the grids (smaller dots). These  r$_{in}$ values cover a range of log \go\ of 1.25 to 4.06 (Section \ref{calculations}). The Cloudy models are run with a source luminosity of 10$^9$ \lsol. Thus, the output masses should be scaled likewise (see Appendix \ref{appendix} for details). The models are run to log N(CO)/cm$^{-2}$ = 17.8  (\Av\ $\sim$ 10 mag) for these grids.}
\label{grids_combine}
\end{figure*}

\subsection{Model Parameters and Variations with Cloud Depth}
\label{grids}
We begin with the grids of \cite{cormier15,cormier19}, who use the spectral synthesis code, Cloudy version 17.00 (Ferland et al. 2017), which simultaneously computes the chemical and thermal structure of \hii\ regions physically adjacent to PDRs.  The central source of the spherical geometry of the Cloudy model is the radiation extracted from Starburst99 \citep{leitherer10} for a continuous starburst of 7 Myr and a total luminosity 10$^9$ \lsol. The grids were computed by varying the initial density at the illuminated face of the \hii\ region, \n, and the distance from the source to the edge of the illuminated \hii\ region, the inner radius, $r_{\rm in}$. The ionization parameter (U) is deduced in the model, based on the input ionizing source, $r_{\rm in}$ and \n. The models are calculated for 5 metallicity bins: Z=0.05, 0.1, 0.25, 0.5 and 1.0 \zsol,\footnote{We assume solar O/H = $4.9 \times 10^{-4}$ \citep{asplund09};  i.e. 12+log(O/H)$_\odot$ = 8.69. } and \n\ ranging from 10 to 10$^4$\,cm$^{-3}$. The r$_{in}$ values range from log(r$_{in}$ cm) = 20.0 to 21.3, in steps of 0.3 dex, which, for the various models, covers a range of U $\sim$ 1 to 10$^{-5}$ and \go\ $\sim$ 17 to 11481 in terms of the Habing radiation field.

 A density profile that is roughly constant in the \hii\ region and increases linearly with the total hydrogen column density (\Nh) beyond $10^{21}$\,cm$^{-2}$ is assumed \citep{cormier19}.
To ensure that all models go deep enough into the molecular phase, regardless of the metallicity, the stopping criterion of the models is set to a CO column density of $10^{17.8}$\,cm$^{-2}$ (\Av\ $\sim$ 10 mag) . With this criterion, the optical depth of the CO(1-0) line, $\tau_{\rm CO}$, is greater than 1 in all models, which, by our definition, means that all models have transitioned in the molecular core\footnote{To apply to specific cases where \Av\ is not necessarily $\sim$ 10 mag, requires running the model to different cloud depths, as we discuss in Section \ref{refinement}, where we quantify the \mhmol\ of the individual DGS galaxies.}. For all calculations, the \Av/\Nh\ ratio is computed self-consistently from the assumed grain-size distribution, grain types, and optical properties. For more description of the Cloudy models that generated the grids analysed here, see Appendix \ref{appendix}.
 
Figure \ref{phase_schematics} shows the evolution of the accumulated mass, abundances and intensities of the \cii, \ci\ and \co\ as a function of metallicity, density, temperature and \go, from the \hplus\ region into the molecular cloud.
As we want to capture the properties of the \hmol\ zone, we have defined the location of the hydrogen ionization front and the \hmol\  front in our calculations. As can be seen in the {\it bottom} subpanels in Fig. \ref{phase_schematics}, the ionization front is defined where H exists in the form of \hplus\ and \hatom\ with a ratio of 50\%/50\%; the transition between \hatom\ and \hmol\ is, similarly, where each species is 50\% of the total hydrogen abundance. Our model results are relatively insensitive to the transition definitions.
The \cplus\  zone is defined to be where 95\% of the total carbon abundance is in the form of \cplus\  and 5\% in the form of  \cone\ and likewise the \cone\ zone has 95\% of the total carbon abundance in the form of  \cone\ (Fig. \ref{phase_schematics} {\it second and third} subpanels). 

To understand the effect of Z, \n\ and \go\ on the zone boundaries and accumulated mass, we compare 3 different cases in Fig. \ref{phase_schematics}:
\begin {enumerate}
\item Metallicity effects:  Compare Z=1.0 \zsol\ ({\it top left} panel) and Z=0.1 \zsol\ ({\it top right} panel) for a fixed \n\ (10$^{3}$\,\den) and fixed \go\ (380); 
\item Density effects: Compare \n\ = $10^2$\, and $10^4$\den\ with a fixed Z (0.5 \zsol) and fixed \go\ (380) ({\it middle left and right} panels); 
\item  \go\ effects: Compare \go\ varying from 80 to 1800 with a fixed Z (0.5 \zsol) and fixed \n\ ($10^3$\den) ({\it bottom left and right} panels). 
\end {enumerate}

\noindent We give here an example to interpret these plots: in the case of Z=1.0 \zsol\ ({\it top left} panel), the transition from \hplus\ to \hatom\ occurs at \Av\ $\sim$ 0.015 mag and the transition from \hatom\ to \hmol\ is at \Av\  $\sim$ 0.5 mag, while the \cplus-emitting zone spans the range of \Av\ $\sim$ 0.05 to 2 mag. 
The evolution of the accumulated mass of the hydrogen species ({\it top left} panel, {\it top} subpanel) indicates that the \cplus-emitting zone includes $\sim$ 90\% of the total gas mass, in the form of both \hatom\ and \hmol, while CO is not yet formed at those \Av\ values. The \cone-emitting region begins at \Av\ $\sim$ 2 mag until CO forms (in a region containing exclusively \hmol).  If we then lower the metallicity to 0.1 \zsol\ ({\it top right} panel), we see that the evolution of the \cplus-emitting zone and the \hatom\ and \hmol\ transitions, in terms of \Av, are the same as the 1.0 \zsol\ case; however the \Nh, or cloud depth, scales differently, requiring a larger cloud depth at 0.1 \zsol\ to reach the same \Av\ as the 1.0 \zsol\ cloud.

The middle panels show how the cumulative \cii, \ci, and \co\ intensities scale when changing only the \n\ in the ionised zone from 10$^{2}$ cm$^{-3}$ ({\it middle left} panel) to 10$^{4} $ cm$^{-3}$ ({\it middle right} panel), while keeping \go\ and Z fixed.  In this case, increasing the \n\ shifts the transitions from \hplus\ to \hatom\ and from \hatom\ to \hmol\ to lower values of \Av. For the higher density case, we are quickly out of the \hplus\ region and into the atomic regime. Also, as can be seen from the {\it center right} panel, \cplus\ is emitting over much of the \hmol\ zone and before CO is emitted for the higher density case.

Finally, the bottom panels demonstrate the effect of changing \go\ and keeping Z and \n\ fixed. The increase of \go\ from 80 to 1800, for example, shifts the \hplus\ to \hatom\ transition to higher \Av\  values (from \Av\ $\sim$ 0.0015 to 0.02 mag) as well as a higher \Nh\ value for the \hatom\ to \hmol\ transition and with \cplus\ tracing a lower mass of \hmol. 

\Av/\Nh\ is computed from the dust-to-gas ratio, assumed to be scaled by metallicity in the model for the range of metallicities studied here.
For an \hplus\ region, as long as dust does not significantly compete with hydrogen for ionizing photons, the \Nh\ of the \hplus\ region will remain constant.  Reducing the metallicity does, however, reduce the \Av\ corresponding to the \Nh\ in the \hplus\ region by the same factor as the metallicity, and hence, the ratio \Av/\Nh.  The size (in \Av) of the \cplus\ zone is roughly independent of metallicity \citep{kaufman06}, since the abundance of \cplus\ scales directly with metallicity, while the UV opacity scales inversely.  For Z=0.1 \zsol\ the carbon abundance is reduced by a factor of ten, but the decrease in dust extinction means that \cplus\ is formed over a path-length ten times longer than that for 1.0 \zsol.  

\subsection{Hydrogen and carbon phase transitions in the model grids}
\label{model_grids}
Another way to visualise the model grids is shown in Fig. \ref{grids_combine} with some observables, such as \hi, \cii, \ci, \ltir\ and \oilinelo. The y-axes of this figure show the \ciitir\ (in percentage) vs \hi\ and \hmol\ gas mass reservoirs in the \cplus\ and \cone-emitting regions (defined above), extracted from Cloudy models for metallicities of Z=0.05, 0.1, 1.0 \zsol\ ({\it from top to bottom}) and for a range of \go\ and \n. The last column  shows the \ciitir\ and \ciioi\ behavior in terms of \go, \n, and Z. 
The {\it total} \hmol\ gas mass from the model
is shown in column 4 for the various metallicity bins. The \cplus-emitting region and some of the \cone-emitting region, depending on the Z, n and \go\ (Fig. \ref{phase_schematics}), will harbour \hmol\ sitting outside the CO-emitting region. Note that the models are scaled to \ltir\ of 10$^{9}$ \lsol. These diagrams can be used to bracket gas masses for a given \ciitir\ and Z bin. To apply these models to observations of a specific galaxy, the masses must be scaled by the proper \ltir\ of the galaxy / 10$^9$ \lsol. More Cloudy model details, including application of the scaling factor, are described in Appendix \ref{appendix}. An example of the application of the model grids to the galaxy, IIZw 40, is demonstrated in Fig. \ref{iizw40} and Section \ref{example}). A simple conversion of observed \cii, \ci\ or \co\ to mass of total \hmol\ or fraction of CO-dark gas is not immediately straightforward. 
 
\subsubsection{Metallicity effects:}
As the metallicity decreases (from bottom to top panels in Fig.\ref{grids_combine}), the grids of gas mass shift to the right: more of the \hmol\ is associated with the \cplus\ and \cone-emitting zones, demonstrating the overall larger CO-dark gas reservoirs predicted by the models in order to reach the molecular core.  The location of both the \hmol\  front and the formation of CO depend on metallicity, with both zones forming at higher \Av\ with decreasing Z \citep{wolfire10}.  The location of the \hmol\ front as a function of \Av\ depends only weakly on metallicity, scaling as $\sim$ $\ln (Z^{-0.75}$).  This is caused by the \hmol\ formation rate scaling linearly with Z, while the UV dissociation rate of \hmol\ depends on both the self-shielding of \hmol\ (independent of Z) and dust extinction (which increases at a given depth with increasing Z).  The CO forming zone scales as $\sim$ $\ln(Z^{-2}$), due to the decreased abundance of oxygen and carbon, which scales directly with Z.  This has the overall effect of increasing the \hmol\ zone.  Additionally, our calculations compute density as a function of column density.  Decreasing Z increases the column density needed to obtain the same \Av\ (also in Fig  \ref{phase_schematics}), therefore making the density in the \hmol\ zone higher at lower metallicity.  These combined effects increase the mass present in the \hmol\ zone.  

\subsubsection{Distribution of the \hatom\ and \hmol\ phases :}
The consequence of metallicity on the partition of the mass between three regions; the \hatom\ associated with \cii\ emission, the \hmol\ associated with \cii\ emission, and the \hmol\ associated with \ci\ emission, can be seen in the first three columns of Fig. \ref{grids_combine}.  The mass of \hatom\ associated with \cii\ increases significantly with decreasing metallicity, as expected. The beginning of the \hatom\ zone starts at the hydrogen ionization front which, for most of our parameter space, is dominated by hydrogen opacity since the dust does not significantly compete with the gas for ionizing photons.  Therefore, the column density corresponding to the hydrogen ionization front is nearly independent of Z, but the \Av\ of the ionization front will decrease with decreasing Z due to the lower \Av/\Nh.   
The \Av\ of the \hmol\ front increases slightly with decreasing Z as a result of decreased dust extinction.  Overall, these two processes lead to a larger \hatom\ zone, and therefore increased mass with decreasing Z.  The size of the \hmol\ zone, and the beginning of the CO-emitting region, follow similar logic.  

\subsubsection{\go\ effects:}
When \go\ increases (for the same density and Z), the \hmol\ and CO fronts are pushed out to higher column densities, as expected, due to an increased dissociation rate.  The width (in \Av) of the \hmol\ zone therefore shows little variation with \go.  Increasing \go\ does have a mild effect on the mass associated with the \hmol\ zone, with increasing \go\ decreasing the mass in the \hmol\ zone (Fig.\ref{grids_combine}).  This is because increasing \go\ moves the \hmol\ zone out to a larger column density (Fig.\ref{phase_schematics}).  Since the density in our models increases with increasing \Nh, the physical size of the \hmol\ zone will shrink.  The thinner physical size leads to a smaller integrated mass, when compared to a lower \go\ calculation.  This effect is small, however, when compared to variations of CO-dark mass with Z.  Increasing density also slightly decreases the \hmol\ region mass, due to slight changes in the location of the \hmol\ front and the beginning of the CO formation zone. 

\subsubsection{Density effects:} 
In addition to the size of each region, the temperature, \n, and \go\ will dictate if the \hatom, \hmol, and \cplus\ regions will emit, and therefore trace the CO-dark gas.  For \cii, the emission is controlled by the C$^{+}$ column density, the $n_{\rm crit}$ for \cii\ emission (3x10$^{3}$ cm$^{-3}$ for collisions with \hatom), and the excitation temperature (92 K) \citep{kaufman99}.  
We see that a density below the $n_{\rm crit}$ in the \hatom\ region allows for efficient emission of \cii\ (Fig. \ref{phase_schematics}).  This is reflected in the fact that, for the lower density models shown in Fig. \ref{grids_combine}, the \hi\ mass and the \hmol\ mass traced by \cii\ are often comparable.  For densities beyond $n_{\rm crit}$, the emission of \cii\ is nearly independent of \n, and is therefore controlled by the temperature and column density.  Since the density increases with column density, for almost all models except the lowest density of 10 cm$^{-3}$, the density in the \hmol\ zone eventually exceeds $n_{\rm crit}$ of \cii, while for the \hatom\ zone the column density is lower, leading to lower densities in this region and a larger region of the parameter space with a density lower than $n_{\rm crit}$ of \cii.  This explains why Fig. \ref{grids_combine} has some spread in the mass for different densities, while for the \hmol, traced by \cii, the plot is nearly constant for decreasing Z, except for the lowest densities considered in our calculations.  For Z = 1 \zsol, a bit more spread is seen since the column density to reach the stopping criterion, and hence the density increase, is smaller. A similar effect occurs for the mass traced by \ci, which has a $n_{\rm crit}$ similar to \cii, thereby causing almost all of the lower metallicity calculations to be independent of \n.  Overall, the reservoir of \mhmol\ where the \cplus\ emission is originating, is systematically lower than that of the \cone-emitting zone. This effect is caused by the fact that where \cone\ is emitting, most of the hydrogen is in the form of \hmol\ while in the \cplus-emitting region, a significant amount of the hydrogen is in the form of \hi, not necessarily in \hmol\ and the thickness of the respective layers are comparable. 
 
The plots of the 5th column in Fig. \ref{grids_combine} show the \cii\ to \oi\ ratio as a function of \ciitir\ for the range of \n\ and \go. \cite{kaufman99} show that this ratio depends strongly on \go\ for low density, especially for the higher Z case. Then for increasing density the functional form of the ratio changes, becoming more sensitive to density, for densities greater than the $n_{\rm crit}$ of \cii.  For larger densities, the \oi\ emission increases with density, while the \cii\ stays roughly constant (for a constant \go), leading to a lower \cii\ to \oi\ ratio.  This explains the trend of this plot with decreasing Z, as the increased density in the PDR, given our density law, leads to models with an initially low density reaching the $n_{\rm crit}$ of \cii, which causes the predicted emission of \oi\ to increase relative to \cii.  

 Figure \ref{grids_combine} allows us to quantify the amount of gas mass accumulated until the stopping criterion of the model is reached (ie log N(CO)=17.8; refer to section \ref{grids}) as a function of important physical parameters, given the observed FIR spectrum.  For example, in the metallicity bin, Z=1.0 \zsol, given an observed \ciitir\ of 0.5\%, the total mass of \hmol\ gas ranges from 1 $\times$ 10$^{8}$ to 3 $\times$ 10$^{9}$ \msol, depending on the density, while for the lowest metallicity bin shown, Z=0.05 \zsol, the quantity of \hmol\ gas ranges from 1 to 3 $\times$ 10$^{10}$ \msol\ for this same \ciitir\ value. This shift to higher mass ranges of \hmol\ gas, suggests that the mass of CO-dark \hmol\ gas can be an important component in low Z galaxies, as already pointed out in previous works \citep[e.g.][]{wolfire10}. 
 
 In summary, from Fig. \ref{grids_combine}, the observed \ciitir\ enables us to determine a range of plausible \go\ and \n\ for a given metallicity.  If \go\ and \n\ can be determined from other assumptions or observations (e.g. \cii/\oi\ for density) then a tighter constraint for the total \hmol\ gas can be determined (section \ref{refinement}), eliminating some of the spread in \n\ and \go. For the lowest metallicity cases, the range of \hmol, determined by \go\ and \n, is relatively narrow, and less dependent on variations in \go\ or \n.  Thus, even having only observations of \cii\ and \ltir, Fig. \ref{grids_combine} may bring a usefully narrow range of solutions for the total \hmol\ for the lowest metallicity cases based on the definition of CO dark gas adopted here and used by \cite{wolfire10}.

\begin{figure*}
\includegraphics[width=9.0cm]{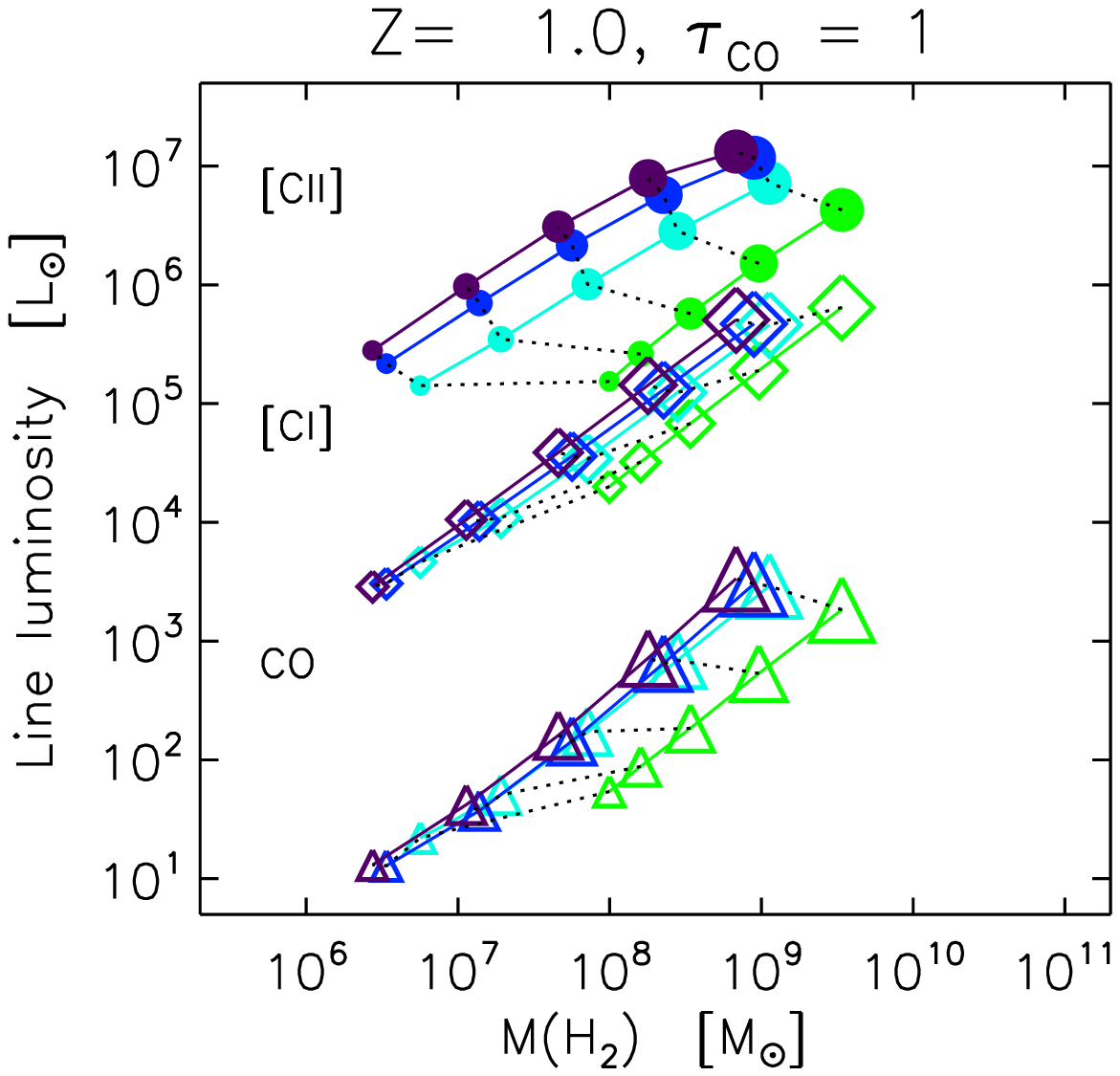}
\hspace{6mm}
\includegraphics[width=9.0cm]{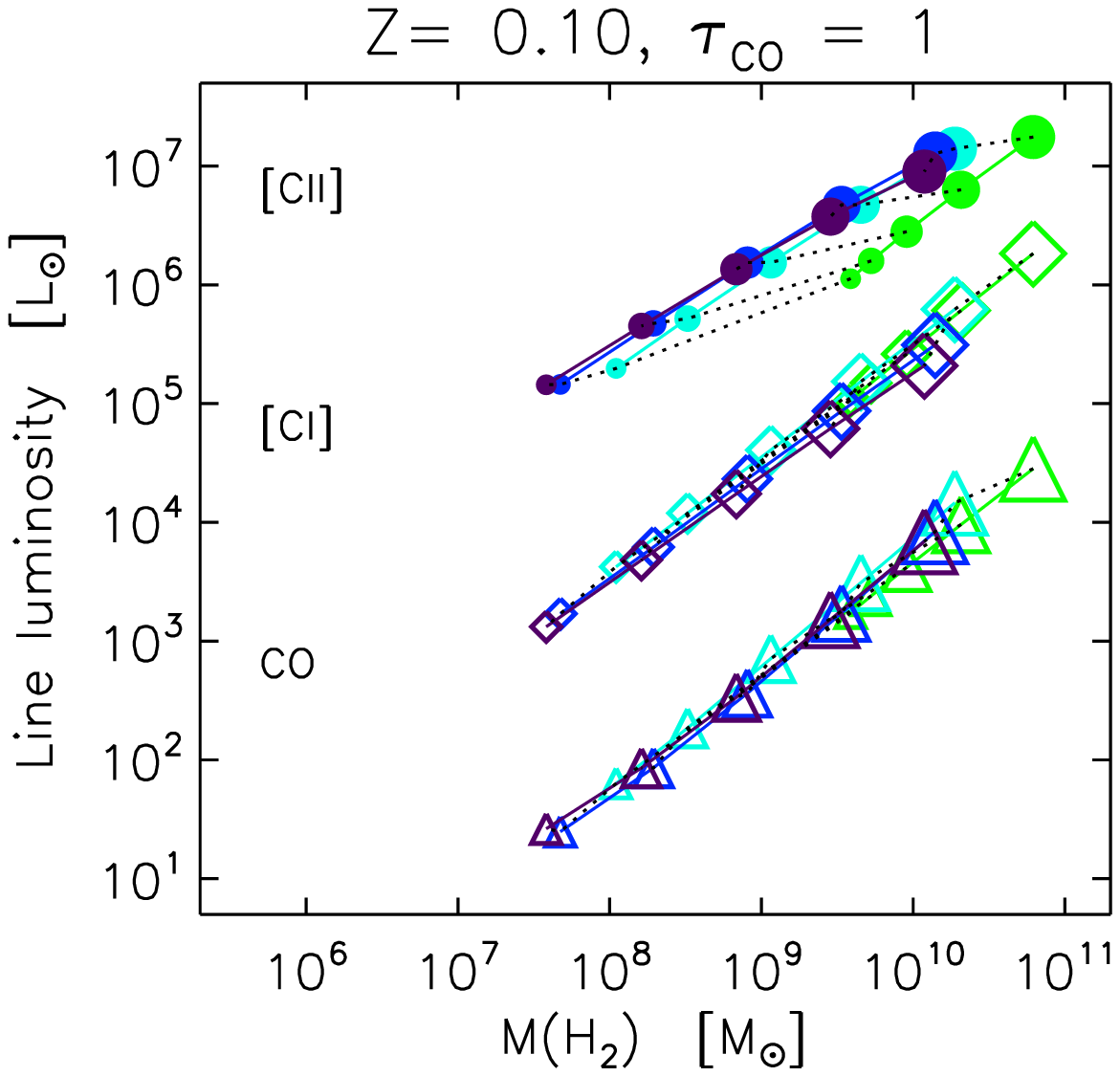}
\caption{The model grids which provide the \lcii, \lci(1-0) (609 $\mu$m) and \lco\ - to \mhmol\ conversion factors for Z=1.0 \zsol\ ({\it left}) and Z=0.1 \zsol\ ({\it right}) for the particular model using a cloud depth of log N(CO)/cm$^{-2}$ = 17.8. The colour coding refers to density values ranging from log \n/cm$^{-3}$ = 1 (green) to 4 (purple). \go\ increases with increasing symbol size: 5, 20, 100, 200 and 500. To obtain the total molecular gas mass, which would include He, an additional factor of 1.36 \citep{asplund09} should also be included for the total mass. Note that \mhmol\  in the CO-emitting region is very sensitive to the depth at which the models are stopped which corresponds to \Av\ $\sim$10 mag here. 
}
\label{line_luminosities}
\end{figure*}
 
\subsection{How \cii, \ci\ and \co\ can trace \mhmol}
We extract the \cii\ and \co\ and  \ci(1-0) (609 $\mu$m) luminosities at the model stopping depth of log N(CO) = 17.8 (\Av\ $\sim$10 mag), inspecting the line luminosities as a function of \mhmol. In Fig. \ref{line_luminosities} we show the \lcii/\mhmol, \lci/\mhmol\ and \lco/\mhmol\ conversion factors as a function of \n\ and \go\ for the examples of Z= 1.0 \zsol\ and Z= 0.1 \zsol. Caution must be exercised when using \lco/\mhmol\ from these figures, since the \mhmol\ within the CO-emitting region, is very sensitive to the depth at which the models are stopped. If there is reason to trust that \Av\ of $\sim$ 10 mag is an accurate representation of the CO-emitting zone, then the \lco/\mhmol\ from this figure would be justifiable. It may not be applicable, for some low Z cases, as discussed in Section \ref{refinement}. 

Note how the profiles shift to the right, to higher \hmol\ masses at lower Z. For the higher Z case, density, in particular, plays an important role in determining the \mhmol\ conversion factors for these species. For low Z, the density variations become less sensitive to the \mhmol, as noted in the Z=0.1 \zsol\ case ({\it right} panel of Fig. \ref{line_luminosities}). For the case of \lci/\mhmol\ and \lco/\mhmol\ at lower metallicity, the density contours collapse together (Fig. \ref{line_luminosities}, {\it right} panel), as is the case for the \lcii, except for the highest \n\ case (\n\ = 10$^4$ cm$^{-3}$; the green dots).  In principle, \ci\ should be able to quantify \mhmol, independent of density. From the right panel of Fig. \ref{line_luminosities}, we derive a \lci/\mhmol\ conversion factor for the case of Z= 0.1 \zsol:

\begin{equation}
\label{ciconversion}
$\mhmol\ = 10$^{4.06}$ $\times$ \lci$^{1.04}$ $
\end{equation}

\noindent where \lci\ is in units of \lsol\ and \mhmol\ is in units of \msol. Considering the small effects of density and \go\ on the \lci/\mhmol\ conversion factor for this particular low Z case, Eq. \ref{ciconversion} determines \mhmol\ from \lci\ with a standard deviation of 0.3 dex.  The \lcii/\mhmol\ is about 2 orders of magnitude higher, while the \lco/\mhmol, 2 orders of magnitude lower than \lci/\mhmol.
\ci\ appears to be a useful tracer to quantify the \mhmol, as also illustrated from hydrodynamical models \cite[e.g.][]{glover16, offner14}, and, as we show here at least for low Z cases, with little dependence on \n. The fact that \lci\ is much fainter than \lcii\ (Fig. \ref{line_luminosities}) makes it more difficult to use as a reliable tracer of \mhmol. These effects can also be seen from the grids in Fig.\ref{grids_combine}.  

We continue in the following sections to demonstrate the use of the observed \cii\ to determine \mhmol\ for specific cases and will followup on \ci\ as a tracer of \mhmol\ in a subsequent publication.
  
\subsection{How to determine \Av\ and its effect on line emission. ("Spaghetti plots")}
\label{refinement}

\begin{figure*}
\includegraphics[width=9.5cm]{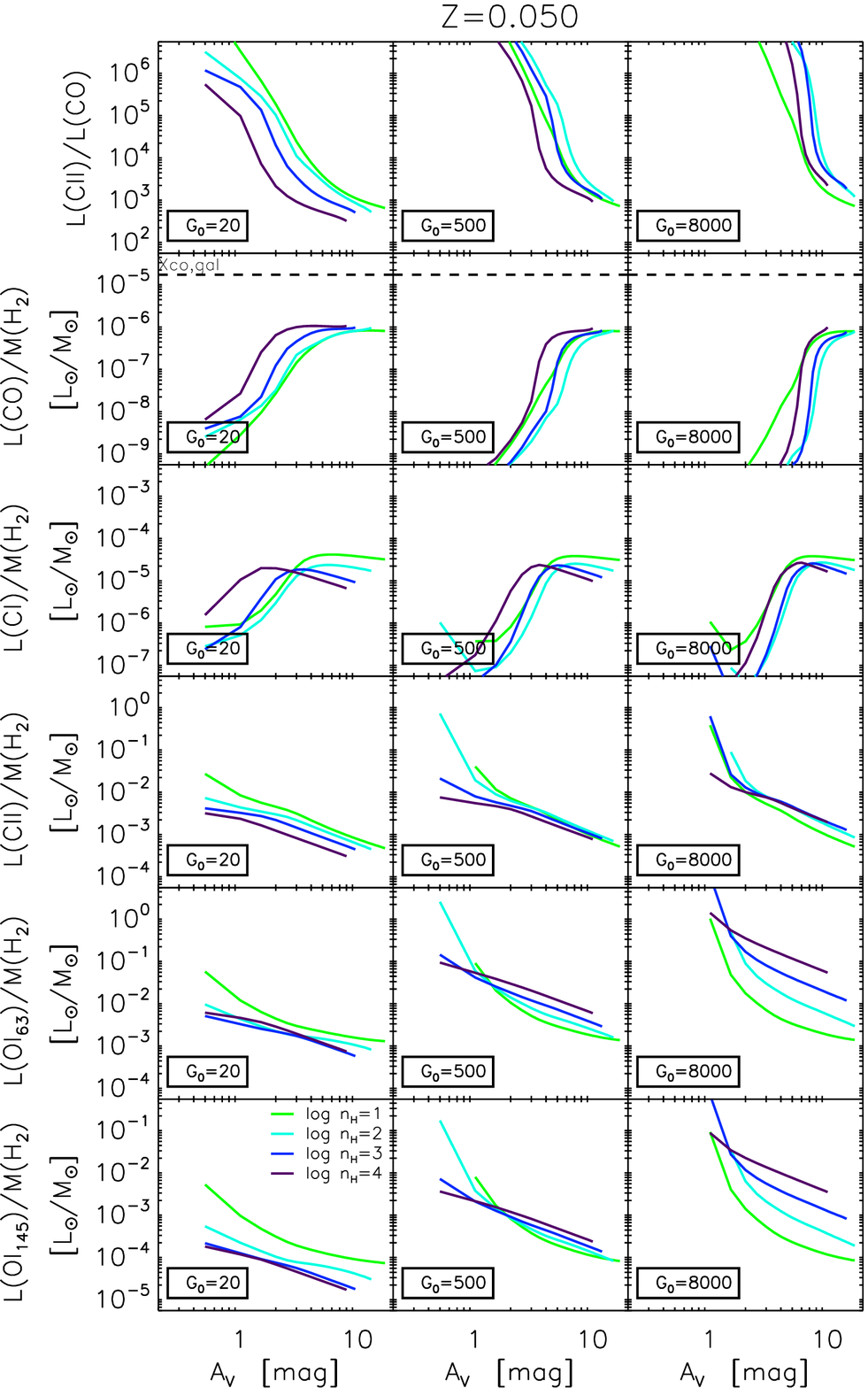}
\includegraphics[width=9.5cm]{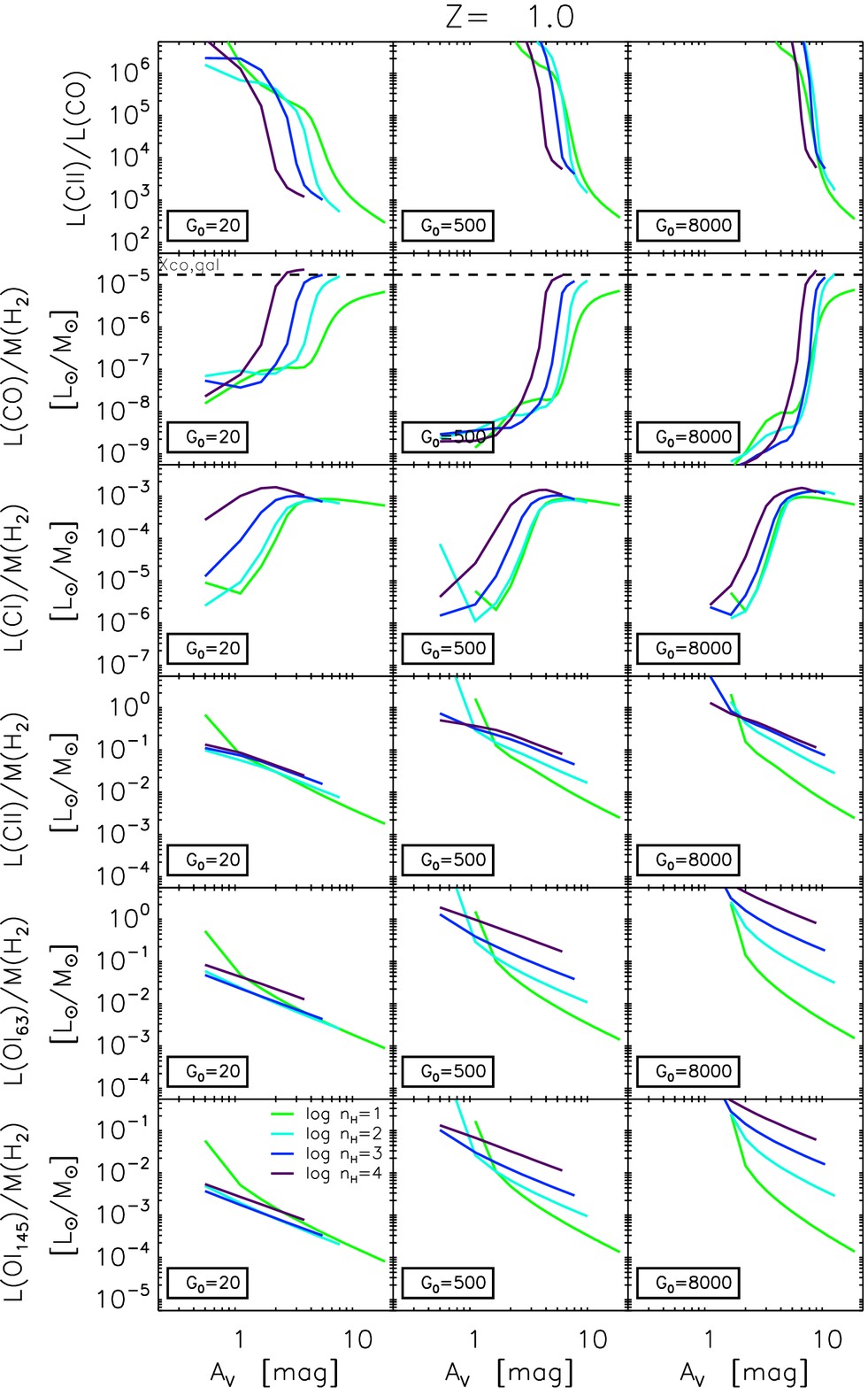}
\caption{"Spaghetti plots": model results for \ciico\ and ratios of \lco/\mhmol, \lci$\lambda$609$\mu$m/\mhmol, 
\loi/\mhmol, as a function of \Av\ for a range of \n\ ($10^{1}, 10^{2}, 10^{3}, 10^{4}$ \den) and \go\ values ($\sim$ 20, 500, 8000) for Z=0.05 \zsol\ ({\it left}) and 1.0 \zsol\ ({\it right}).}                                                                                                                                                                
\label{spaghetti}
\end{figure*}

We see from Fig. \ref{cii_co_fir_fig} that  \ciico\ exhibits great variations from galaxy to galaxy. When comparing the observed \ciico\ for low metallicity galaxies (ranging from $\sim$  {3\,000 to 80\,000), to the modeled \ciico, where the stopping criterion is  \Av\ $\sim$ 10 mag, we see that the models do not reach such high observed \ciico\ values. This is because when the models are stopped at \Av\ $\sim$ 10 mag, too much CO has already formed for the low Z cases (Fig. \ref{phase_schematics}), which require the models to stop at lower \Av. The higher metallicity molecular clouds, exhibiting lower \ciico\ than the dwarf galaxies, could be better described by stopping at \Av\ $\sim$ 10 mag. To explore the sensitivity of the emission as a function of depth, we can use the observed \ciico\ as a powerful constraint. 

Figure \ref{spaghetti} shows the model grid results for the \ciico\ and ratios of \cii, \co, \ci$\lambda$609$\mu$m, \oilinelo\ and \oilineup\ lines to \mhmol\ as a function of \Av\ for a range of \n\ ($10^{1}, 10^{2}, 10^{3}, 10^{4}$ \den) and \go\ values ($\sim$ 20, 500, 800) for Z=0.05 \zsol\ (\it left column}) and Z=1.0 \zsol\ ({\it right column}).  We can understand the behavior of these plots also referring to Fig. \ref{phase_schematics} and Sections \ref{grids} and \ref{model_grids}.

For example, when \Av\ $\sim$ 1 mag is reached, \cii\ formation is increasing rather linearly while CO increases exponentially (also evident in Figure \ref{phase_schematics}), and between \Av\  $\sim$ 1 and 10 mag we see a rather linear decrease of \ciico.  The \lcii/\mhmol\  and \loi/\mhmol\ keep decreasing beyond \Av\ $\sim$ 1 mag since the \cii\ and \oi\ have stopped emitting but the \mhmol\ continues to increase, so the overall trend is the drop in \lcii/\mhmol\  and  \loi/\mhmol\ for greater \Av. Density can have a considerable effect on the  \oi\ emission, producing a wide spread of \loi/\mhmol\ as \Av\ increases, for higher \go\ environments. For example, for \go\ of 8000 and \Av\ $\sim$ 5, there is $\sim$ 50 to 100 times greater \loi/\mhmol\  for \n\ =10$^4$ cm$^{-3}$, compared to \n\ = 10$^1$ cm$^{-3}$ for Z = 0.5 and 1.0 \zsol. Both \oi\ lines behave similarly, with \oilineup\ generally weaker than \oilinelo\ by $\sim$ one order of magnitude.

The \ciico\ is a good tracer of \Av\ for all \go\ and \n, with the range of \Av\ becoming narrower for the spread of \n\ as \go\ increases, as seen in the top panels of Figure \ref{spaghetti}.  The \lci/\mhmol\ begins to turn over, peaking at \Av\ of $\sim$ a few mag, more or less where the \cii\ formation has decreased. This is why we see a relative flattening of the \lci/\mhmol\ for growing \Av.

We stop the plots in the figures beyond where CO has formed and becomes optically thick. Otherwise this \mhmol\ will continue to accumulate causing the \lco/\mhmol\ to turn over and begin to decrease beyond the point of $\tau_{\rm{CO}}$ = 1. Having \co\ observations in addition to the \cii\ observations, brings the best constraint on the \Av\ of the cloud and thus a better quantification of the \mhmol.
                                                                 
Note that to construct Fig. \ref{spaghetti}, the models are stopped at log N(CO) = 17.8 (a maximum \Av\ $\sim$10 mag) and the line intensities are extracted at the different depths into the cloud - at different cloud \Av\ values. In principle, the emerging line intensities could be different depending on the stopping criterion, due to optical depth effects and cloud temperature structure. 
To quantified this effect, we ran grids stopping at several maximum \Av\ values (2, 5 and 10, for example) and extracted line intensities, comparing the values we present in Fig. \ref{spaghetti}. The effects on the \oi\ and \cii\ line intensities are mostly negligible (variations $<  20\%$ throughout the grid).  The intensities of the grid stopping at a maximum of \Av\ of 2 mag are generally similar to or larger than the line intensities extracted at the cloud depth of \Av\ = 2 mag of a grid stopping at maximum \Av\ of 10 mag, primarily due to optical depth effects on the grids run to \Av\ of 10 mag versus \Av\ of 2 mag. 
The \co\ can see up to a factor of 2.5 variations (primarily at \Av\ = 2 mag), depending on the stopping criteria and the extraction of line intensities in \Av\ construction.  Our comparison verification has assured us that we can move forward with our use of Fig. \ref{spaghetti}.

Later, when the models are applied to particular observations (Section \ref{example}), the depth of the models, i.e. the maximum \Av, will be adapted for the specific metallicity case, to determine the range of associated masses in the different zones. More precise determination of the stopping criterion will also require some knowledge of the range of \go\ and \n\ (for example, \ciioi, in Fig. \ref{grids_combine}). Only then can we have the necessary ingredients to obtain the total \mhmol.  
Thus, determination of \Av\ then gives us the total \mhmol.The mass of CO-dark \hmol\ is then the difference between the total \mhmol\ from the models and the \hmol\ determined from the observed \co\ using a Galactic \xco.

\subsection{Quantify the total \mhmol\ and CO-dark gas - how to use the models}
\label{use_models}
We walk through the steps to constrain the total \mhmol\  and the CO-dark gas, depending on
the availability of observations. We emphasise that to obtain the total molecular gas, including He, an additional factor of 1.36 \citep{asplund09} should be taken into account. The best-constrained case is that for which \cii\xspace, \oi\xspace, \co\xspace and \lfir\xspace
observations exist. We also illustrate the range of solutions that can
be obtained with less observational constraints.

Since metallicity plays an important role in the models, knowledge or assumptions of Z is necessary. The parameters that define
the applicable model are \go\xspace, \n\xspace, and the maximum
\Av\xspace. We outline the steps to use the models:
\begin{itemize}
\item  First we use the \ciifir\xspace to find the range of \go\xspace using Fig.~\ref{grids_combine}\footnote{We provide the values of these plots in table form at the CDS where the reader can select model values of Z, \n, inner radius (r$_{in}$), \go, \Av\ and $\tau_{CO}$ to obtain predicted \mhmol, \cii, \co, \ci\ and \oi\ luminosities}. As can been seen in those figures the \ciifir\xspace is most strongly
dependent on the radiation field density and less so on density for a given Z. 
\item The second step is to further refine the
corresponding model parameters with the \ciioi\xspace ratio
(right-most panels in Fig.~\ref{grids_combine}), or any combination of tracers that would provide the density in the \cii-emitting zone of the PDR. For a given (range
of) radiation field(s) this ratio is sensitive to the gas density.
The combination of these two observational ratios generally constrains
the radiation field and starting density well.
\item Finally, the depth of the model (\Av\xspace) can be found using the
observed \ciico\xspace in combination with Fig.~\ref{spaghetti}. The
top series of panels shows the predicted line ratio for the
different combinations of Z, \go\xspace and \n\xspace.
\end{itemize}
\noindent These steps are illustrated for one galaxy, as an example, in Section \ref{example} and Fig. \ref{iizw40}.

Having found the parameter combination(s) that reproduce the relative
strength of these key emission lines and the dust continuum, we can
scale the model to the absolute line strength using the panels above
and obtain the mass of \hmol\xspace. Probably the most useful scaling
is based on using the \lcii/\mhmol\ predictions (4th
row of each panel in Fig.~\ref{spaghetti}) because the \cii\xspace line is strong and it
originates throughout the parts of the model where the \ciico\xspace
varies strongly. Note that our methodology using \ci\xspace as an
observational constraint, does not (directly) aid in better determining
the most applicable models in the range of interest, i.e. those
situations where the CO-emitting zone is reached. However, the
conversion from \ci\xspace line luminosity to molecular gas mass for
the applicable models is tighter than for \cii\xspace (see also
Fig.~\ref{line_luminosities}). Therefore \ci\xspace observations will
be very useful for getting the best possible measures of the total
\hmol\ gas mass once the matching models have been determined and compared to observations. 
Once the total \mhmol\ is determined, the difference between this value and the \mhmol\ determined from \co\ and the \xco\ conversion factor, will quantify the CO-dark gas reservoir. 
 
\section{Applying the models: the particular example of \iizw}
\label{example}
 Here, we use one galaxy from the DGS, \iizw\xspace, (Z=0.5~\zsol\xspace), to
demonstrate how to determine the total \mhmol\ directly from the observations and the models presented above and thus the subsequent CO-dark gas mass. The relevant steps
are visualised in Fig.~\ref{iizw40}, where the grids are run to a maximum depth of \Av\ = 10 mag. 
We first obtain the fiducial model results for \mhmol, using the basic set of observational constraints: \cii, \oi, \co\ and \lfir. Often it is not possible to have all of these tracers. Thus,  we also explore the derived ranges of \mhmol\xspace for \iizw\xspace when we have limited observational data available to constrain \Av\xspace, \n\xspace and \go\xspace. In this way we can get an idea of what kind of accuracy can be obtained, depending on the available constraints.

\subsection{Fiducial model}
\label{fiducial_model}
\begin{figure*}
\includegraphics[width=14.0cm]{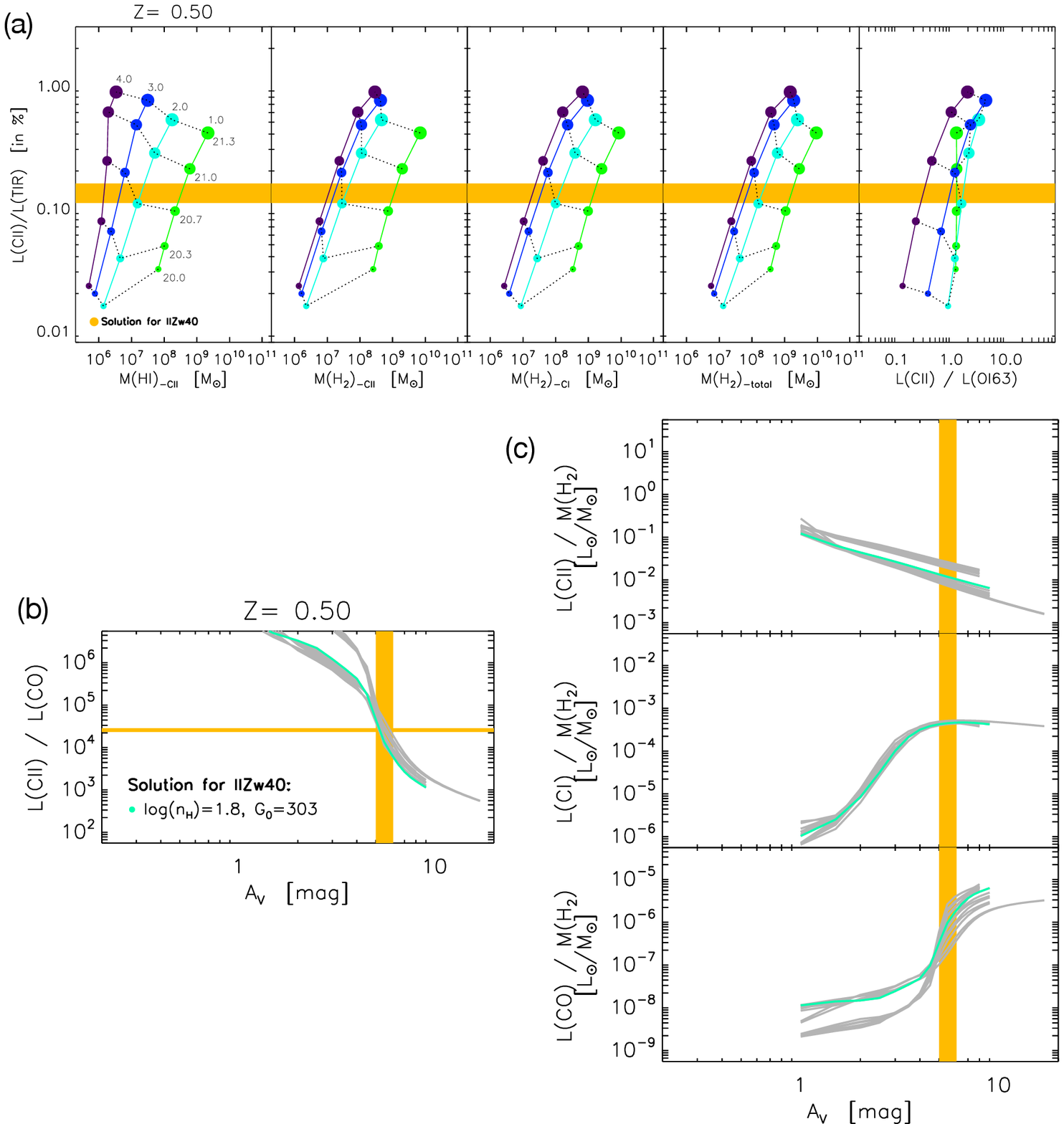}
\caption{Applying our model to the particular case of II Zw 40. Panels (a) show how the observed \ciitir\ and \ciioi\ line ratios allow us to constrain \n\ and \go. Panel (b) shows how the \ciico\ ratio evolves in \iizw\ as a function of \Av, given the parameters previously derived from Panel (a). The green line indicates the best matching model. Panels (c) finally show the corresponding factors to convert \lcii, \lci\ and \lco\ to \mhmol (see Section \ref{example}).}
\label{iizw40}
\end{figure*}

The observed value of \ciitir\xspace in \iizw\xspace is
0.134$\pm$0.024 \citep{cormier15} which translates to \go\xspace around 300 for the range of
densities considered. The \ciitir\xspace alone does not tightly constrain the density. In this case we also have the valuable \oi\ line. The relatively high value of \ciioi\xspace
(1.35$\pm$0.29) matches the lower density models (log(\n\xspace)=
1.8, from interpolation) (last panel of Fig.~\ref{iizw40}a).

We retain the models that match the combined \ciitir\xspace and \ciioi\xspace, i.e.
the models that cross the yellow intersection in the right most panel
of Fig.~\ref{iizw40}a. Figure~\ref{iizw40}b, extracted from Fig.~\ref{spaghetti}, shows the behavior of
\ciico\xspace as a function of \Av\xspace for these models in grey. The green line
is the model curve of the best matching model. As can be seen, all
models reproduce the observed \ciico\xspace (2.33\, $\times$ 10$^{5}$) at an \Av\xspace
value of $\sim$5 mag with a small dispersion.

Finally, the top panel of Fig.~\ref{iizw40}c shows that the
\lcii\xspace/\mhmol\xspace  for the best matching model depth
is $\sim$0.01 [\lsol/\msol] but values up to $\sim$0.02 are also
compatible. The \lcii\xspace of 3.87\, $\times$ 10$^6$ \lsol\xspace of
\iizw\xspace translates to a total \mhmol\xspace ranging from
1.3$-$5.1\, $\times$ 10$^8$\,\msol\xspace with the best matching model yielding
a \mhmol\xspace of 3.2\,$\times$ 10$^8$\,\msol\xspace.

It is interesting to compare the derived \hmol\ gas mass with that 
determined using \lco\xspace and a standard (Milky Way) \xco\xspace conversion factor
\footnote{In this study for the standard \xco\xspace conversion we use a factor of 2\, $\times$ 10$^{20}$ cm$^{-2}$(K km s$^{-1}$)$^{-1}$ in terms of \xco\ (I$_{CO}$ $\propto$ N$_{H2}$);
3.2 \msol\ pc$^{-2}$ (K km s$^{-1}$)$^{-1}$ in terms of $\alpha_{\rm CO}$ (I$_{\rm CO}$ $\propto$ \mhmol). While $\alpha_{\rm CO}$ is normally 4.3 \msol\ pc$^{-2}$ (K km s$^{-1}$)$^{-1}$, including helium, here we do not include the helium mass (factor of 1.36) when comparing to the model output of \mhmol.}. The standard conversion factor yields a value of 7.1\, $\times$ 10$^6$ \msol\xspace, i.e. only a few percent of the actual total \hmol\ gas mass determined from these models. We already know that as Z decreases, \xco\ increases and calibrations for \xco\ based on metallicity vary wildly in the literature \citep[e.g.][]{bolatto13}. Thus, comparing with a Galactic \xco\ really gives a lower limit on what is expected. In any case, even for the moderately low Z of 0.5 \zsol, the CO-dark molecular gas will be important.

\subsection{Estimating \mhmol\xspace with fewer observational constraints}
Following the full example for \iizw\xspace above, we consider: case {\it a}) when
using \cii\xspace, \co\xspace and \ltir\xspace (no \oi\ line observed or no other density indicator) and case {\it b}) only \cii\xspace and \ltir\xspace are available (\co\ and \oi\ are not available). In each case, we compare to the \mhmol\ derived from our fiducial model above (Section \ref{fiducial_model}), where a more complete set of observations (\cii, \ltir, \oi\ and \co) was available.
\begin{itemize}
\item case {\it a}: Using \cii, \ltir\xspace and \co. The observed ratio of
  \ciitir\xspace for this galaxy (Fig. \ref{iizw40}) indicates combinations
  of log(\n\xspace) and log(\go\xspace), from (1;2.2) through
  (2.5;2.8) to (4.0;3.3). For these models \Av\xspace values between 3.5 and
  6 mag reproduce \ciico\xspace within its uncertainty. The best model, i.e.
  the closest predictions to the observed values, has an \Av\xspace of 4.5 mag
  and contains $1.3\, \times 10^8$\,\msol\xspace of \hmol\xspace. The \hmol\ gas
  mass in the models that satisfactorily reproduces the observations,
  ranges from 0.6 to 5.1\, $\times$ 10$^{8}$\,\msol\xspace. Comparing with the range we
  find for the fiducial model (1.3$-$5.1\, $\times$ 10$^8$\,\msol\xspace), 
  we can see that the allowed range is significantly larger: the upper value is the same, but now allows a lower end of the range. In the particular case of \iizw\xspace the higher density models that contain less \mhmol\ gas before reaching the observed \ciico\xspace values can not be excluded and the range is thus expanded to lower values.

\item case {\it b}: Using \cii\xspace and \ltir\xspace as the only available observational constraints; \co\xspace has not been observed
  or with limited sensitivity and \oi\xspace or another density tracer, has not been observed. This means that we cannot exclude
  ``normal'' \ciico\xspace ratios and high optical depth. Without a
  constraint on the \Av\xspace, these models can only be used to infer an
  upper limit on the \mhmol in the PDR by integrating the model until
  $\tau_{CO}$ $\simeq$1. The same combinations of \n\xspace and \go\xspace as for
  case {\it a} yield upper limits on \hmol\ gas masses of 0.7 to
  13\, $\times$ 10$^{8}$\,\msol\xspace. Removing the CO-bright \hmol\ gas from these
  models by using the predicted \lco\xspace with a Galactic conversion
  factor yields upper limits on the dark \hmol\ gas mass of
  0.5 to $11\, \times$ 10$^{8}$\,\msol. Thus the derived upper limit of
  13\, $\times$ 10$^{8}$\,\msol\xspace is significantly above the value we derive
  (3.2$\pm$1.9\, $\times$ 10$^8$\,\msol\xspace) using the full set of constraints.
  Whether such an upper limit is useful will depend on the specific
  science question one is trying to address.
\end{itemize}
 
\section{Quantify the total \mhmol\ and CO-dark gas in the Dwarf Galaxy Survey}
\label{dgs} 
 To quantify the CO-dark gas of the DGS galaxies, we determine the predicted total \mhmol\ for each DGS galaxy from the models (\mhmoltot).  We then compare this predicted total \mhmol\ to the mass of the CO-bright \hmol, \mhmolco, using the observed \lco\ and Galactic \xco. The mass of the CO-dark gas component, \mhmolnoco, would then be the difference between the model-predicted \mhmoltot\ reservoir and the observed \mhmolco:

\begin{equation}
\label{mass}
$\mhmolnoco\ = \mhmoltot\ - \mhmolco\ $
\label{total}
\end{equation}

We note \n\ values ranging from $10^{0.5} - 10^3$\,\den\ and \go\ values of $10^2-10^3$ that are determined from the \cite{cormier19} model solutions for the DGS galaxies, for metallicities ranging from near solar to $\approx 1/50$ \zsol \footnote{Metallicity values for the DGS galaxies are from \cite{madden13} which uses the strong line metallcity calibration from \cite{pilyugin05}. }
. We then extract the total \mhmol\ for each galaxy from the corresponding model grids, applying the steps described above, to quantify the total \mhmol\ consistent with the model solutions and, consequently, derive the CO-dark gas reservoir for each galaxy. Various galactic properties, observed and modeled parameters, and their relationships with the total \hmol\ mass and CO-dark gas reservoirs, are inspected (Figs. \ref{dgsplots1}, \ref{dgsplots2} and \ref{dgsplots3}). 

\subsection{Trends of \mhmoltot\ and CO-dark gas with model parameters}
\begin{figure*}
\includegraphics[width=18.0cm]{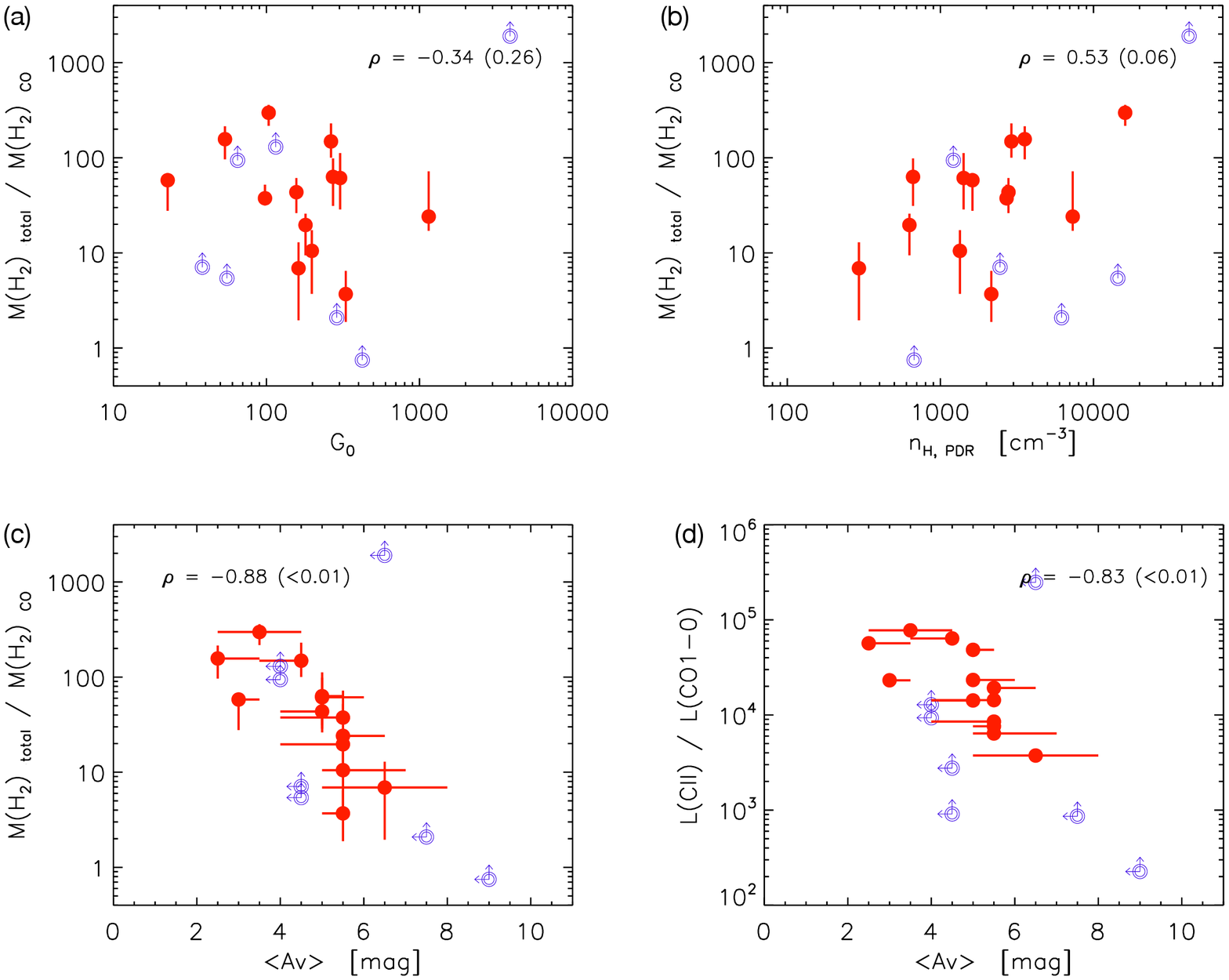}
 \caption{Results from applying the  to the DGS sample and trends with model parameters. The vertical axis in panels a) through c) is \mhmoltot\ / \mhmolco\ (the ratio of the total \mhmol\ determined from the model and the CO-bright \mhmol\ determined from CO observations and the Galactic conversion factor) versus on the horizontal axis:  (a) \go; (b) density and (c) \Av.  (d) Observed \ciico\ vs. model \Av. Spearman correlation coefficients ($\rho$)  and p-values in parenthesis are indicated within each panel. Red points are solutions for the DGS galaxies with CO detections. Open symbols are upper limits due to CO non-detections. The masses determined by the model for the individual galaxies have been scaled by their proper \ltir.  
}                                                                                                                                                                                                                                                                
\label{dgsplots1}
\end{figure*}

We notice right away (Fig. \ref{dgsplots1}) that for DGS galaxies, \mhmoltot\ is always much larger than that determined using CO only, \mhmolco. The  increasing \mhmoltot\ / \mhmolco\ signals the effect of the CO-dark gas reservoir becoming an increasingly important component of the total \mhmol, particularly in low Z environments,  which has been noted in the literature \citep[e.g.][]{poglitsch95,israel96,madden97,wolfire10, fahrion17,nordon16, accurso17a,jameson18,lebouteiller19}.  The total mass of \hmol\ can range from 5 to a few hundred times the \hmol\ determined from the CO-emitting phase.
{\it The CO-dark gas dominates the total \hmol\ reservoir in these galaxies. We are clearly missing the bulk of the \hmol\ by observing only CO.}  What is controlling the fraction of CO-dark gas? 

The \go\ and density determined from the models do not seem to play an obvious role in driving the CO-dark gas fraction as shown in Fig. \ref{dgsplots1} panels a and b. Panel c shows the tight anti-correlation of \mhmoltot\ / \mhmolco\ with \Av, and, as shown in panel d, \Av\ anti-correlates with \ciico, underscoring the role of \Av\ in regulating the \cplus\ - CO phase transition in the PDR  \cite[e.g.][]{wolfire10,nordon16,jameson18}.  The extreme range of observed \ciico\ in low metallicity galaxies seen in Fig. \ref{cii_co_fir_fig} is a consequence of their overall low average effective \Av.

\subsection{Trends of \mhmoltot\ and CO-dark gas with observables}
\begin{figure*}
\includegraphics[width=18.0cm]{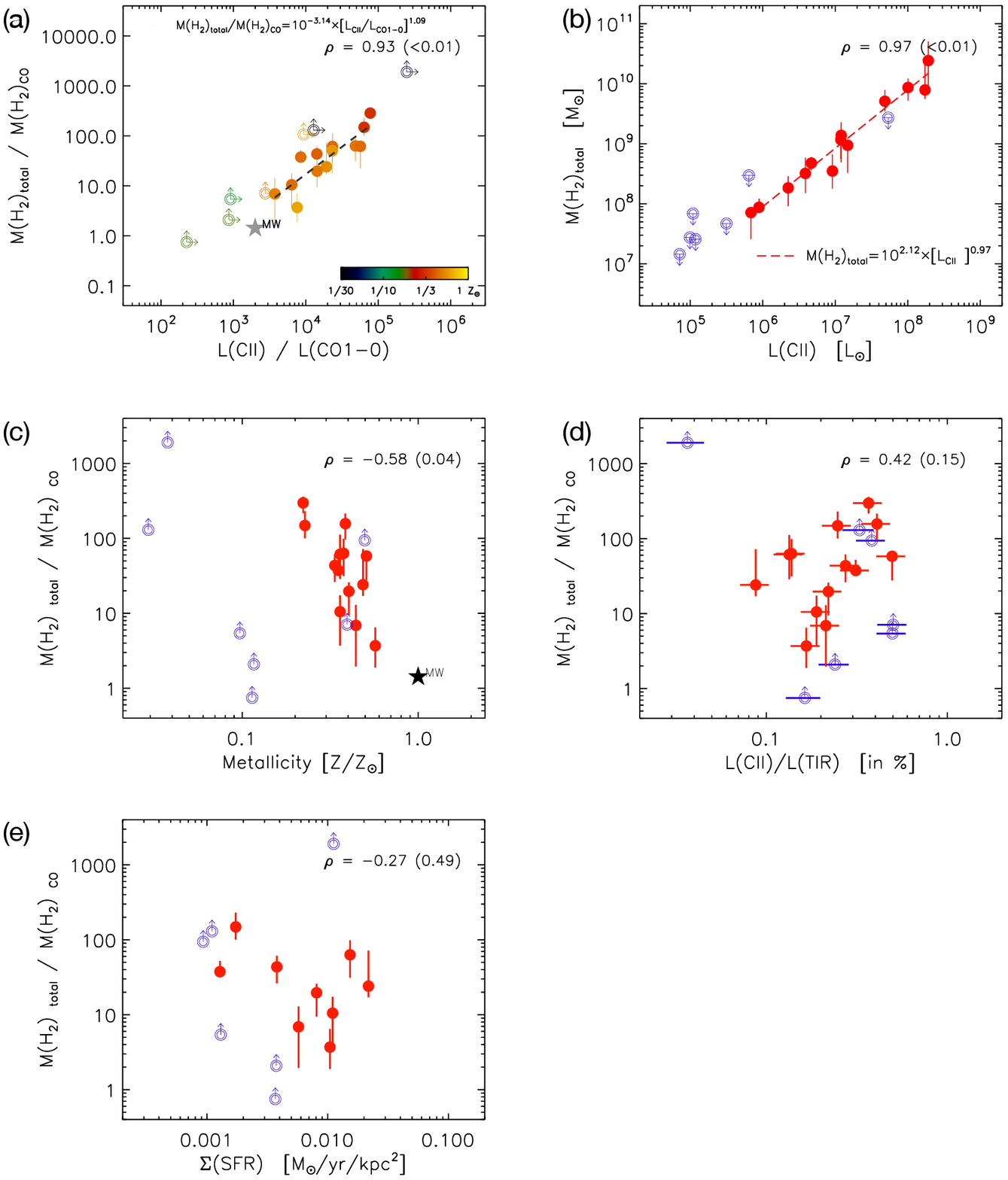}
 \caption{Results for the DGS sample and trends with observable parameters. The vertical axis of a), c), d), e): \mhmoltot\ / M(\hmol)$_{\rm CO}$ (total \mhmol\ determined from the model over the \mhmol\ determined from CO observations and the Galactic conversion factor) versus:
\newline
(a) \ciico\ with colour code for Z. The correlation shown in the dashed line is described in the panel in the equation for \mhmoltot\ / M(\hmol)$_{\rm CO}$ as a function of the observed \ciico; standard deviation is 0.25 dex.
\newline
(b) total \mhmol\ determined from the models versus the observed \lcii.  Our resulting relationship of \mhmol\ as a function of \lcii\ is given within the panel; standard deviation is 0.14 dex.
\newline
(c) \mhmoltot\ / M(\hmol)$_{\rm CO}$ versus Z; 
\newline
(d) \mhmoltot\ / M(\hmol)$_{\rm CO}$ versus observed \ciitir\ and 
\newline
(e) \mhmoltot\ / M(\hmol)$_{\rm CO}$ versus $\Sigma_{\rm{SFR}}$, where SFR is determined from total infrared luminosity from \cite{remy15}; 
\newline
\newline
Spearman correlation coefficients ($\rho$)  and p-values in parenthesis are indicated within each panel. 
Red points are DGS galaxies with CO detections. Open symbols in all panels are upper limits due to CO non-detections. 
}                                                                                                                                                                                                                                                         
\label{dgsplots2}
\end{figure*}

What relationships exist between the observed quantities or measured galaxy properties and the total \mhmol\ and the quantity of CO-dark gas? In Fig. \ref{dgsplots2} (panel a) we see that the observed \ciico\ is an excellent tracer of the CO-dark \mhmol\ fraction. We fit this correlation to convert from observed \ciico\ to the \mhmoltot/\mhmolco: 

\begin{equation}
$\mhmoltot\ / \mhmolco\ = 10$^{-3.14} \times\ $ [\ciico]$^{1.09}$$
\label{eq3_ciico}
\end{equation}

\noindent The standard deviation of this fit is 0.25 dex. It follows, from eq. \ref{total} and eq. \ref{eq3_ciico}, that the ratio of the mass of CO-dark gas to CO-bright gas is, therefore:

\begin{equation}
$\mhmolnoco\ / \mhmolco\ = 10$^{-3.14} \times\ $ [\ciico]$^{1.09}$ - 1.0$
\label{eq4_ciico}
\end{equation}

We find a very tight correlation between \lcii\ and total \hmol\ gas mass (Fig. \ref{dgsplots2}, panel b). The observed \lcii\ can thus convert directly to {\it total} \mhmol:

\begin{equation}
\label{eq5_ciico}
\label{ciimhmol}
$\mhmoltot\ = 10$^{2.12} \times\ $[\lcii]$^{0.97}$$
\end{equation}

\noindent with a standard deviation of 0.14 dex.
The \cplus\ luminosity alone can pin down the total \mhmol, making the \ciiline\ a valuable tracer to quantify the molecular gas mass in galaxies. Our modelling results find about a factor of 3 larger mass of total \hmol\ associated with \lcii, and hence a larger reservoir of CO-dark gas mass, than that determined empirically from \cite{zanella18}. \cite{zanella18} determine the \mhmol\ from an assumed CO-to-\mhmol\ conversion factor which is lower then that found in this study. We find a trend of the CO-dark \mhmol\ fraction growing as the metallicity decreases (Fig. \ref{dgsplots2} panel c). 

 Our study of the lowest-metallicity objects is, however, limited by the lack of robust CO detections in these extreme environments, thus limiting our knowledge of the behavior of Z with the CO-dark gas mass or total \mhmol\ at the lowest Z end. The CO-dark gas mass fraction does climb steeply as the Z decreases, even for moderately low-Z galaxies. 

We see a weak trend of increasing fraction of CO-dark gas mass as the observed \ciitir\ increases (Fig. \ref{dgsplots2} panel d). For the relatively narrow range of \ciitir\ there is a wide range of CO-dark gas fraction. The fraction of CO-dark gas, which covers almost 2 orders of magnitude, does not show a trend with the $\sim$ 1 order of magnitude range of surface density of SFR ($\Sigma_{\rm{SFR}}$) in our galaxy sample (Fig. \ref{dgsplots2} panel e). 
This is consistent with the effect of increasing \ciitir\ (larger fraction of CO dark gas) being correlated with the increasing \lcii\ (Fig. \ref{dgsplots2} panel d) and less so with direct effects of \ltir.  

\subsection{Consequences of the CO-dark gas fraction on the Schmidt-Kennicutt relation and the X$_{CO}$ conversion factor}
\begin{figure*}
 \includegraphics[width=18.0cm]{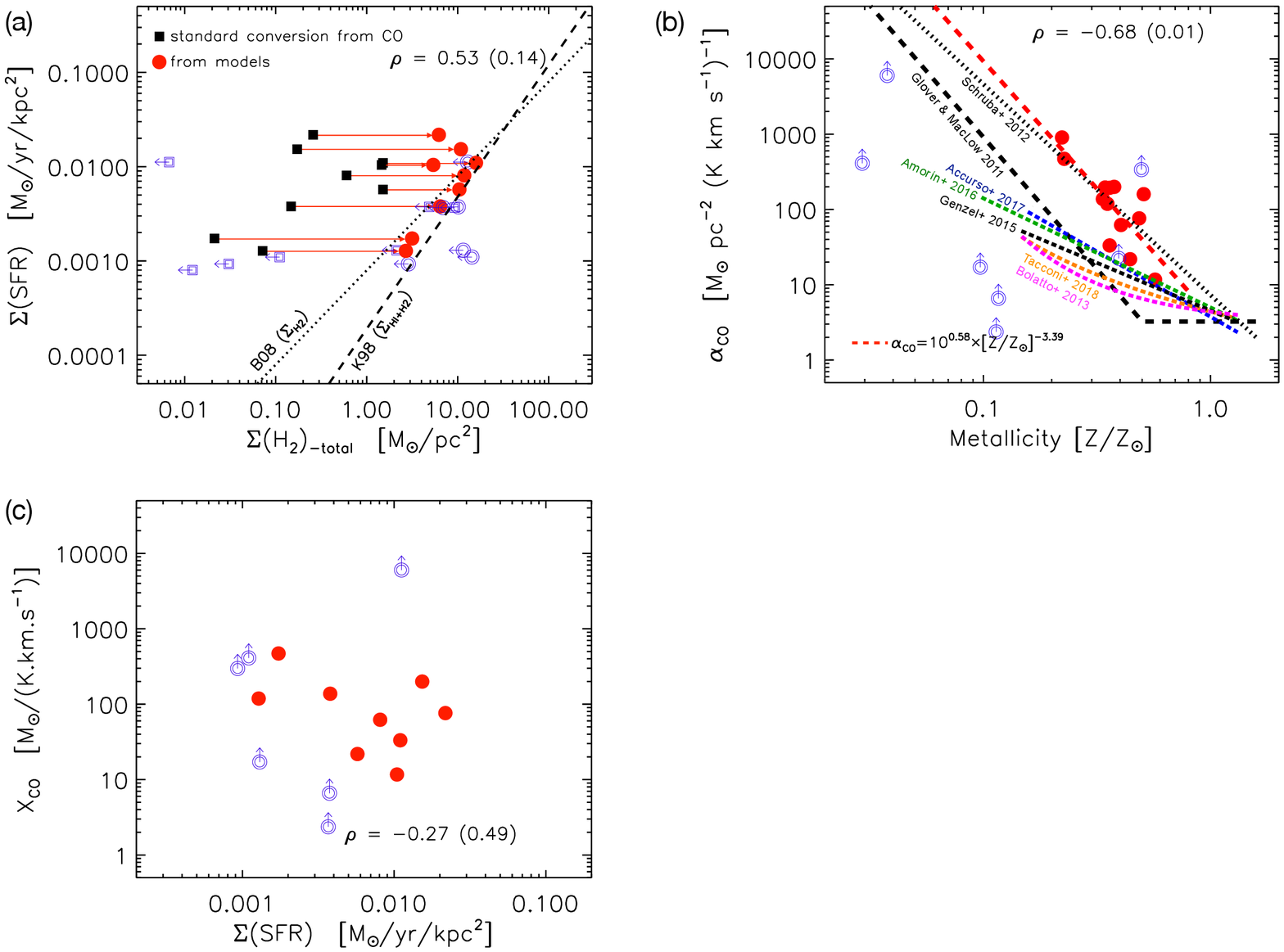}
\caption{Consequences of quantifying the {\it total} \hmol\ of the DGS sample.
\newline
(a) The Schmidt-Kennicutt relationship and Bigiel (2008) revisited. Solid black squares are the values when the \mhmol\ is calculated from the CO-to-\hmol\ standard conversion factor; solid red dots are the  {\it total} \hmol\ determined from our self-consistent models. The dashed line is the usual Schmidt-Kennicutt relation, where low-Z star-forming dwarf galaxies are normally outliers when CO is used to determine the \hmol\ \citep{cormier14}. The dotted line is the Bigiel (2008) relationship determined for the  $\Sigma_{\rm{SFR}}$-$\Sigma_{\rm{gas}}$ relationship. 
\newline
(b)  $\alpha_{\rm CO}$ as a function of Z from \cite{schruba12} (black dotted line),  \cite{glover11} (long black dashed line), \cite{accurso17b} (blue dashed line), \cite{amorin16} (green dashed line), \cite{genzel15} (short black dashed line), \cite{tacconi18} (orange dashed line), \cite{bolatto13} (pink dashed line) and our new $\alpha_{\rm CO}$ -  Z relationship determined from this paper (red dashed line). Red solid dots are the  {\it total} \hmol\ determined from our self-consistent models in this paper. Also given in the panel is the derived expression to determine $\alpha_{\rm CO}$ as a function of Z from this new relationship, which has a standard deviation of 0.32 dex. 
\newline
(c) X$_{CO}$ conversion factor from this paper and $\Sigma_{\rm{SFR}}$. Spearman correlation coefficients ($\rho$) and p-values in parenthesis are indicated within each panel.Red dots are DGS galaxies with CO detections. Open symbols in all panels are upper limits due to CO non-detections.
}                                                                                                                                                                                                                                                         
\label{dgsplots3}
\end{figure*}

What is the consequence of the presence of this reservoir of CO-dark \hmol\ on $\Sigma_{\rm{SFR}}$ and the surface density of gas ($\Sigma_{\rm{gas}}$) in galaxies, as described in the relationships of \cite{kennicutt98b} and \cite{bigiel08}? In  Fig. \ref{dgsplots3} (panel a), we determine the $\Sigma_{\rm{M_{H_2}}}$ within the CO-emitting region (black squares) and find their positions well-off the  $\Sigma_{\rm{M_{H_2}}}$ - $\Sigma_{\rm{SFR}}$ relationships, as found from \cite{cormier14}, which may be suggesting much higher $\Sigma_{\rm{SFR}}$ for their \mhmol. Once we take into account the total \mhmol\ determined from \cii\ and the modelling, which now includes the CO-dark \mhmol\ as well as the CO-bright \mhmol\ (Fig. \ref{dgsplots3}, panel a; red dots), we see the locations of the galaxies shift to the right, lying between both $\Sigma_{\rm{M_{H_2}}}$ - $\Sigma_{\rm{SFR}}$ relationships. The CO-dark gas is an important component to take into account in understanding the star formation activity in dwarf galaxies. While the data are limited, we find that taking into account the {\it total} \mhmol, the star-forming dwarf galaxies do show a similar relation as the star-forming disk galaxies. Thus they are not necessarily more efficient in forming stars.

It has been shown from simulations \citep[e.g.][]{glover12,krumholz11b} that star formation can proceed without a CO prerequisite as well as without \hmol. The relationship observed between \hmol\ and star formation can be due to the cloud conditions providing the ability to shield themselves from the UV radiation field, thereby, allowing them to cool and to form stars. In this process of shielding, at least \hmol\ formation can also proceed, as well as CO in well-shielded environments. The conditions required to reach the necessary low gas temperatures set the stage for efficient \cplus\ cooling. CO may accompany star formation but does not have a causality effect. Thus it may not come as a surprise to see star-forming dwarf galaxies, harboring a dearth of CO, showing a similar relationship as that of the more metal-rich disk galaxies seen in the Schmidt-Kennicutt relationship.

With our determination of total \hmol\ we can now give an analytic expression to convert from observed CO to a total \mhmol\ conversion factor,  $\alpha_{\rm CO}$,  and its relationship with Z. Here, again, the total \mhmol\ now includes the CO-dark gas plus the CO-bright \hmol\ mass. We find from Fig. \ref{dgsplots3} (panel b):

\begin{equation}
\label{alphaco}
\alpha_{\rm CO} = 10^{0.58} \times [Z/$\zsol$]^{-3.39}
\end{equation}

\noindent with a standard deviation of 0.32 dex. We find a steeply rising $\alpha_{\rm CO}$ as the metallicity decreases. For example, at Z= 0.2 \zsol, $\alpha_{\rm CO}$ is about 2 orders of magnitude greater than that for the Galaxy.  A strong dependence of the CO-to-\hmol\ conversion factor on metallicity is not unexpected. Since the survival of molecules depends on how unsuccessful UV photons are in penetrating molecular clouds and photodissociating the molecules, extinction plays an important role in this process. Thus, the lower dust abundance in the low metallicity cases, comes into play.  In Fig. \ref{dgsplots3} (panel b) we show comparison of our derived metallicity dependence of $\alpha_{\rm CO}$ with others in the literature. \cite{schruba12} have determined $\alpha_{\rm CO}$ from the observed SFR scaled by the observed \lco\ and a constant depletion time. This approach assumes that the efficiency of conversion of \hmol\ into stars is constant within different environments. Our determination of $\alpha_{\rm CO}$ and that of \cite{schruba12}\footnote{While different relations have been shown for different categories of galaxies in \cite{schruba12}, we use here the relation given for all of the galaxies in that study}, are quite comparable, given the spread of the observations, low number statistics, as well as possible uncertainties in metallicity calibrations. The $\alpha_{\rm CO}$ from our study seems to climb even more steeply toward lower Z. However, the lower end of the metallicity space, where mostly only upper limits in \co\ observations exist, is not pinned down robustly.  It is well shifted up from the \cite{glover11} relationship with CO-to-\mhmol\ conversion factor, which is determined from hydrodynamical simulations. A shallower slope is found for star-forming low metallicity galaxies from \cite{amorin16} who derive a metallicity-dependent $\alpha_{\rm CO}$ considering the empirical correlations of SFR, CO depletion time scale and metallicity. Other $\alpha_{\rm CO}$-Z scaling functions, such as \cite{arimoto96} and \cite{wolfire10} fall near that of \cite{amorin16}.

By studying how the molecular gas depletion times vary with red-shift and their relation to the star-forming main sequence, \cite{genzel15} have determined a scaling of $\alpha_{\rm CO}$ taking into account CO and dust-based observations (Fig. \ref{dgsplots3}, panel b). While this conversion factor is shallower to that we find from our study of local low-Z dwarf galaxies, they note that their study, which is based on massive star-forming galaxies, is probably not reliable for Z $<$ 0.5 \zsol. \cite{accurso17a} determine a comparable scaling of $\alpha_{\rm CO}$ with Z as  \cite{genzel15} using similar surveys,  but includes a second order dependence on distance from the star-forming main sequence in their $\alpha_{\rm CO}$. They caution against using this relation below Z$\sim$ 0.1 \zsol, where it has not been calibrated. \cite{bolatto13} take a thorough look at numerous observations and theory and different metallicity regimes and propose a Z-dependent $\alpha_{\rm CO}$ which is similar to \cite{genzel15} for near-solar Z galaxies, but curves to steeper $\alpha_{\rm CO}$ for lower Z cases, while \cite{tacconi18} propose a compromise between \cite{bolatto13} and \cite{genzel15} (Fig. \ref{dgsplots3}, panel b). 

As soon as the ISM of the galaxy is more metal-poor, given the observed CO which, even for moderately metal-poor galaxies, is difficult to obtain, the conversion factor from CO-to-\hmol\ quickly grows. Even at 20\% \zsol, the CO conversion factor is already 1000 times that of our Galaxy. The true reservoir of \mhmol\  may have been severely underestimated so far at low Z and these new relations can quantify that.

\section{Possible caveats and limitations}
While \cii\ emission can be a convenient tool to quantify a reservoir of molecular gas that is not traced by \co, there are some caveats and limitations to this study.
\begin{itemize}
\item Lower metallicity bound: We emphasize that these relationships to determine total gas mass presented here have only been studied for the star-forming dwarf galaxies of the DGS with metallicities as low as $\approx{1/50}$ \zsol. The models have been applied for the low metallicity galaxies for which CO is observed and this has limited the derived $\alpha_{\rm CO}$ and \lcii-to-\mhmol\ conversion factor only to metallicities as low as Z $\sim$ 0.2 \zsol, even though \cii\ has been detected in DGS galaxies below this metallicity.
\item Limited range of model parameters:  The models have been applied over a range of  \n\ = 10 to 10$^4$ \den\ and over a range of inner radii, corresponding to log \go\ $\sim$ 1 to 4. These results need to be studied over broader ranges of galactic properties to be applied with confidence beyond this study.
For example, faint, low metallicity, more quiescent dwarf galaxies can also habor star formation rates lower than those found for the Schmidt-Kennicutt relation \cite[e.g.][]{roychowdhury09,cigan16}, in contrast to the DGS sample (Fig. \ref{dgsplots3} panel a). Some of these more quiescent dwarf galaxies are at the lowest metallicity range explored here and do not have CO detections, but may also be harboring some CO-dark molecular gas. Likewise, more massive, CO-rich galaxies have yet to be tested with this model.
\item Galactic size scales: These conclusions have been drawn on unresolved galaxy scales. The reliability of \cii\ to trace \mhmol\ on resolved scales using these models has not been tested. A similar approach was carried out at 10 pc resolution in the 30Doradus region of the low metallicity LMC (Z = 0.5 \zsol), where $>$ 75\% of the \hmol\ was CO-dark and traced by \cii\ \citep{chevance20}. Considering that the \cii\ emission is expected to be more extended than the \co\ \cite[e.g.][]{jameson18,chevance20}, our simple  scheme using Cloudy and representing galaxies with a single \hii\ region + molecular cloud in one dimension, does not account for realistic geometry and for the physical mixing of different physical components and thus, mixing \go\ and \n\ conditions. 
\item The origin of \cii\  emission in galaxies: Note that \cii\ can arise from other components in galaxies, not only from neutral PDRs. For example \cii\ can originate in the low density ionised gas component in galaxies, excited by electrons. In principle, to use \cii\ to quantify the \mhmol, it is necessary to first determine if there is "contamination" of the \cii\ emission arising from the ionised gas  which should first be removed before application of the model. This requires observations of other ionised gas tracers, such as the commonly-used FIR \nii\ lines, for example. Such studies often conclude that, most of the \cii\ emission in galaxies arises from PDR regions, with a decreasing fraction of the total \cii\ emission arising from the ionised gas with decreasing metallicity \citep{kaufman06,croxall17,jameson18,cormier19, herrera_camus16,accurso17b,sutter19}. Also in our study we have not taken into account any contributing molecular gas reservoirs originating in the WNM or CNM atomic phases. The atomic gas component in our Galaxy has been proposed to harbor almost 50\% of dark molecular gas of \mhmol\ \cite[e.g.][]{kalberla20}.  We do extract the \hi\ mass from the models, but with the observations of our global studies, we can not yet explore this component. Most of the \hi\ in the dwarf galaxies arises throughout the galaxies and often from a very extended component - extending well beyond the \cii\ emission. We do not resolve the \hi\ associated with the PDRs, and thus, can not address this point.
\item Metallicity calibrations: A word of caution related to metallicity determination is necessary when applying these scaling relations. Metallicity calibrations are well-known to differ \citep[e.g.][]{kewley08} and where absolute calibrations differ significantly, this should be taken into account when looking into detailed application of the models.
\end{itemize}

While this study points to the usefulness of \cii\ as a tracer of the molecular gas in low-Z galaxies, enlarging the galactic parameter space will be needed to apply these findings in a more general sense to a wide variety of galaxies, especially galaxies more massive than the dwarf galaxies in this study. Here we only focus on galaxies of the local universe. However, this is already a potential step forward in the possibility of estimating the gas mass from high-z galaxies using ALMA to access the \ciiline, in the cases where \co\ may be faint, perhaps due to lower metallicity.

\label{caveats}

\section{Summary and Conclusions}
This study is motivated by the extreme \ciico\ values observed for low metallicity galaxies - almost reaching 10$^{5}$, on global scales, which can be up to an order of magnitude higher than dustier star-forming galaxies. The bright \ciiline\ lines observed in low metallicity galaxies have been challenging to reconcile with the faint or undetectable \co\ in light of their star formation activity. We have investigated the effects of metallicity, gas density and radiation field on the total molecular gas reservoir in galaxies and quantified the mass of \hmol\ not traced by \co\ - the  CO-dark molecular gas which can be traced by \cii. Cloudy grids traversing these physical parameters are inspected to understand the behavior of observed quantities, such as \oi, \ci, \co, \cii\ and \lfir\ in terms of \mhmol, as a function of \Av, metallicity and \n.  In principle, \ci\ can be an important tracer of the CO-dark molecular gas, and this will be further investigated in a follow-up study.  However, due to its higher luminosity, \cii\ is an ideal tracer of the molecular gas. We give recipes on how to use these models to go from observations to total \mhmol. We apply the models to the \hers\ Dwarf Galaxy Survey, extracting the total \mhmol\ for each galaxy. The CO-dark \mhmol\ is then determined from the difference between the total \mhmol\ and the CO-bright \hmol, traced by the observed \co. Our findings indicate that \co\ in the dwarf galaxies traces only a small fraction of the \hmol, if any, while, the {\it total \mhmol\ is dominated by the CO-dark gas which can be uncovered by \cii\ observations}. 
\newline

\noindent We have determined a \lcii-to-\mhmol\ conversion factor:
\noindent \mhmoltot\ = 10$^{2.12} \times\ $[\lcii]$^{0.97}$   
\noindent with a standard deviation of 0.14 dex.
Following from this, we give a new CO-to-\mhmol\ conversion factor that takes into account the total \mhmol\  - both the CO-dark and CO-bright gas - given by application of the models:

\noindent $\alpha_{\rm CO}$ = 10$^{0.58} \times [Z/$\zsol$]^{-3.39}$ 
with a standard deviation of 0.32 dex.
\newline

\noindent Comparisons with the fraction of CO-dark gas in the low metallicity galaxies and their galactic properties reveal these findings:

\begin{enumerate}
\item The fraction of CO-dark gas is correlated with \ciico. There is a tight correlation between the \ciiline\ and the total \mhmol\ over the range of low metallicity galaxies of the DGS.
\item The effective \Av\ from our models is anticorrelated with \ciico\ and hence the CO-dark gas fraction. Hence the consequence of the effective low \Av\ overall, in low metallicity galaxies is the extreme \ciico\ observed in star-forming dwarf galaxies.
\item The SFR, \n\ and \go\ do not individually control the CO-dark gas mass fraction.
\item The CO-dark gas accounts for most ($>$ 70\%) of the total \hmol, over the wide range of galaxy properties of the \hers\ Dwarf Galaxy Survey. This study consists of galaxies with \ciitir\ ranging between 0.1 to 0.5\%, metallicity values from $\sim$ 0.02 to 0.6 \zsol, \ciico\ values corresponding to 2 orders of magnitude of \mhmol\ and over 3 orders of magnitude of \lfir\ and stellar mass. 
\item Our new \co-to-\mhmol\ conversion factor as a function of metallicity, which also accounts for the CO-dark \mhmol, is steeper than most others in the literature, but resembles that of \cite{schruba12}.
\item When taking into account this significant CO-dark \hmol\ reservoir, the star-forming dwarf galaxies, which were above the Schmidt-Kennicutt relation when the observed \co\ only was used to deduce the \hmol\ reservoir, now shift to the normal relation of $\Sigma_{\rm{SFR}}$ - $\Sigma_{\rm{M_{H_2}}}$ found for disk, star-forming galaxies.
\item \ci\ shows signs of being a convenient tracer of the \mhmol, particularly at low Z, where it can be independent of \n\ and \go. However, \ci\ is intrinsically less luminous than \cii. The utility of \ci\ as a tracer of total \mhmol\ for a variety of galactic conditions, requires further follow-up investigation.
 \end{enumerate}

We conclude that the low \lco/SFR and the high \ciico\ observed in low metallicity dwarf galaxies can be explained by the photodissociation of CO, and signals the presence of a prominent reservoir of CO-dark \hmol. Observations of a larger number of extremely low-Z galaxies is necessary to pin down the lowest Z end of the $\alpha_{\rm CO}$ - Z and the \lcii\ - \mhmol\ relation as well as expanding the studies of parameter space and spatial scales in galaxies. Likewise, robust comparison of other methods to determine \mhmol\ and when these different methods are applicable, will be the subject of followup studies.

\label{conclusions}

\begin{acknowledgements}
The authors wish to thank the anonymous referee for a number of suggestions that have improved the presentation of this study. We thank A. Poglitsch for valuable discussions that also helped to improve the quality of the paper. We acknowledge support from the DAAD/PROCOPE projects 57210883/35265PE  and from the Programme National {\it Physique et Chimie du Milieu Interstellaire (PCMI)} of CNRS/INSU with INC/INP co-funded by CEA and CNES. The European UnionÕs Horizon 2020 research and innovation program under the Marie Sklodowska-Curie grant agreement No 702622 supported DC during this study. SH acknowledges financial support from DFG programme HO 5475/2-1. MC acknowledges funding from the Deutsche Forschungsgemeinschaft (DFG, German Research Foundation) through an Emmy Noether Research Group (grant number KR4801/1-1) and the DFG Sachbeihilfe (grant number KR4801/2-1). MG has received funding from the European Research Council (ERC) under the 397 European Union Horizon 2020 research and innovation programme (MagneticYSOs project, grant agreement 398 No 679937). FLP acknowledges funding from the ANR grant LYRICS (ANR-16-CE31- 0011). This research was originally made possible through the financial support of the Agence Nationale de la Recherche (ANR) through the programme SYMPATICO (Program Blanc Projet NR-11-BS56-0023) and through the EU FP7.

This study made use of \hers\ observations: PACS was developed by MPE (Germany); UVIE (Austria); KU Leuven, CSL, IMEC (Belgium); CEA, LAM (France); MPIA (Germany); INAFIFSI/OAA/OAP/OAT, LENS, SISSA (Italy); IAC (Spain). This development was supported by BMVIT (Austria), ESA-PRODEX (Belgium), CEA/CNES (France), DLR (Germany), ASI/INAF (Italy), and CICYT/MCYT (Spain). SPIRE was developed by Cardiff University (UK); Univ. Lethbridge (Canada); NAOC (China); CEA, LAM (France); IFSI, Univ. Padua (Italy); IAC (Spain); SNSB (Sweden); Imperial College London, RAL, UCL-MSSL, UKATC, Univ. Sussex (UK) and Caltech, JPL, NHSC, Univ. Colorado (USA). This development was supported by CSA (Canada); NAOC (China); CEA, CNES, CNRS (France); ASI (Italy); MCINN (Spain); Stockholm Observatory (Sweden); STFC (UK); and NASA (USA).
\end{acknowledgements}

\bibliographystyle{aa}
\bibliography{/Users/madden/Latex/references}

\appendix
\section{Cloudy model details}
\label{appendix}
For this study we use Cloudy models similar to those originating from \cite{cormier19,cormier15} for which we have adopted a closed spherical geometry with an internal and isotropic ionizing source surrounded by a 4$\pi$ steradian layer of \hii\ region, and a neutral PDR layer surrounding the ionised layer. The source ionises the inner edge of the cloud and the radiation is propagated through the \hii\ region, self-consistently making the transition into the PDR. The evolution of mass components is calculated based on integrating the density over the volume of the spherical shell at each depth into the cloud. 
 
The model central source is a continuous starburst of 7 Myr from Starburst99 \citep{leitherer10} for a total source luminosity of 10$^9$ \lsol. Cloudy allows the choice of calculations in luminosity or in intensity. We selected the luminosity case for the SED brightness, where the photon luminosity of the central source impacts the cloud beginning at the inner edge of the ionised region (at the start of the \hii\ region), $r_{\rm in}$, which, for these models is varied from log($r_{\rm in}$ cm) = 20.0 to 21.3, in steps of 0.3 dex. The initial hydrogen density (\n) at the illuminated edge of the cloud is varied from 10 to 10$^4$ cm$^{-3}$. While the initial ionization parameter, U, can be an input command only in the intensity case, for our luminosity case, we chose to effectively have a range of U in the luminosity case, by varying  $r_{\rm in}$ as an input parameter, which in turn varies U, which is deduced within Cloudy for each model:

\begin{equation}
U = \frac{Q({\rm H)}}{4\pi r_{\rm in}^2 n_{\rm H} {\rm c}}
\label{U}
\end{equation}

\noindent where Q(H) is the number of hydrogen-ionizing photons emitted by the central source, and $c$ is the speed of light. The intensity of the FUV radiation at the PDR front, \go,  deduced in our Cloudy models, ranges from $\sim$ 17 to 11481 in units of the Habing radiation field (1.6 $\times$ 10$^{-3}$ ergs cm$^{-2}$ s$^{-1}$)\citep{habing68}. Five metallicity bins were calculated for the models: Z=0.05, 0.1, 0.25, 0.5 and 1.0 \zsol.

The dust and PAH properties used in this model are described in \cite{cormier19}. The opacity curves of the SMC are used. The abundance of PAHs is further reduced by metallicity to the power of 1.3, characterizing the prominent drop in PAH abundance at lower metallicity \citep[e.g.][]{remy14,remy15}. Tests performed in \cite{cormier19}, inspecting the sensitivity of the PAH abundance, concluded that the PAH abundance is not an important factor in the outcome of the model results since the grain abundance is always larger than the already-reduced PAH abundance.

In Cloudy it is possible to choose a constant density throughout the cloud or to assume that the total pressure is constant. \cite{cormier19} have looked into different approaches of the density law, for a constant pressure case versus a constant density case and have found a compromise of these 2 extreme cases that works best in predicting the observations for the DGS sources. An intermediate case for the assumed density profile is constructed which is constant in the \hii\ region and increases linearly with the hydrogen column density in the neutral gas.  
The different cases of constant density, constant pressure and smoothly-increasing density law were tested in \cite{cormier19} where, for example, the constant density models predict less \oi\ emission and the constant pressure models predict more \oi\ emission than the smoothly-increasing density law adopted, which was found to successfully describe the observations.

A full galaxy has numerous stellar clusters and \hii\ regions and ensembles of PDR/molecular clouds. To go from a model single cluster-plus-cloud system to a representation of a full galaxy with numerous   cluster-plus-cloud systems, we create a "unit model" with total source luminosity = 10$^{9}$ \lsol. 
The "unit model" of 10$^9$ \lsol\ chosen here is arbitrary but serves as a representative order of magnitude of luminosity of the observed galaxies. Thus all of the line luminosities predicted by the model correspond to a model for which L = 10$^{9}$ \lsol. The output masses are also determined for a model with source L = 10$^{9}$ \lsol. Since mass scales with luminosity, the masses determined for an individual galaxy, having \ltir(galaxy), are scaled by \ltir(galaxy) / 10$^{9}$ \lsol. This approximation assumes an energy balance between UV-optical and infrared and that \ltir\ scales similarly as the luminosity in the hydrogen-ionizing energy range \citep{cormier12}.

We refer to \cite{cormier19} for further details and input parameter studies of a similar model used for this study.
 
 \end{document}